\documentclass{ar-1col}
\usepackage{url}
\usepackage[numbers]{natbib}
\usepackage[cmex10]{amsmath}
\usepackage{comment}

\jname{Annu. Rev. Biomed. Eng.}
\jvol{25}
\jyear{2023}
\doi{10.1146/((please add article doi)). This is a pre-print version of the article which has been accepted for Publication in Volume 25 of the Annual Review of Biomedical Engineering (2023). Only Personal Use is Permitted.}

\begin{document}

\markboth{Chatterjee et al. $\bullet$ Bio-sensors for IoB}{Bio-sensors for IoB}

\title{Bioelectronic Sensor Nodes for Internet of Bodies}

\author{Baibhab Chatterjee,$^{1,2}$ Pedram Mohseni,$^3$ and Shreyas Sen$^1$
\affil{$^1$Elmore Family School of Electrical and Computer Engineering, Purdue University, West Lafayette, IN 47907 USA and Center for Internet of Bodies (C-IoB), Purdue University, West Lafayette, IN 47907 USA;\\email: bchatte@purdue.edu, shreyas@purdue.edu}
\affil{$^2$Department of Electrical and Computer Engineering, University of Florida, Gainesville, FL 32611 USA; email: chatterjee.b@ufl.edu}
\affil{$^3$Department of Electrical, Computer and Systems Engineering, Case Western Reserve University, Cleveland, OH 44106 USA and Institute for Smart, Secure, and Connected Systems (ISSACS), Case Western Reserve University, Cleveland, OH 44106 USA; email: pedram.mohseni@case.edu}
}

\begin{abstract}
Energy-efficient sensing with Physically-secure communication for bio-sensors on, around and within the Human Body is a major area of research today for development of low-cost healthcare, enabling continuous monitoring and/or secure, perpetual operation. These devices, when used as a network of nodes form the Internet of Bodies (IoB), which poses certain challenges including stringent resource constraints (power/area/computation/memory), simultaneous sensing and communication, and security vulnerabilities as evidenced by the DHS and FDA advisories. One other major challenge is to find an efficient on-body energy harvesting method to support the sensing, communication, and security sub-modules. Due to the limitations in the harvested amount of energy, we require reduction of energy consumed per unit information, making the use of in-sensor analytics/processing imperative. In this paper, we review the challenges and opportunities in low-power sensing, processing and communication, with possible powering modalities for future bio-sensor nodes. Specifically, we analyze, compare and contrast (a) different sensing mechanisms such as voltage/current domain vs time-domain, (b) low-power, secure communication modalities including wireless techniques and human-body communication, and (c) different powering techniques for both wearable devices and implants.
\end{abstract}

\maketitle


\tableofcontents

\section{INTRODUCTION}
\label{introduction}

In today's data driven world, modern medical devices and bio-sensor nodes benefit from seamless connectivity, capturing valuable patient-specific information which is subsequently analyzed to gain insights through artificial intelligence (AI), enabling closed loop bio-electronic medicine such as smart insulin pumps, connected pacemakers, neuro-stimulators and performance monitoring devices, among many others. However, the connectivity to these devices through traditional wireless techniques typically results in (a) high communication power, and (b) vulnerability to hacking as the wireless signals can be picked up by a nearby eavesdropper \cite{Das_SciRep_2019}. Evidently, the communication power consumed in a sensor node is usually orders of magnitude higher than the sensing and computation power \cite{Chatterjee_DnT_2019, Cao_VLSI_2020}. However, a sensor node on, in or around the human body does not need to communicate continuously due to data redundancy either in temporal, or in spatial domain. Spatio-temporal In-sensor Analytics (ISA), in the form of compressive sensing, anomaly detection and collaborative intelligence has recently been shown to reduce the communication power by orders of magnitude, while losing $<$2\% information \cite{Chatterjee_JIoT_2021}. ISA reduces the transmission payload of the sensor, thereby reducing the communication power, which in turn, reduces the amount of power needed to sense and transmit unit amount of information. 

While ISA reduces the energy requirements of sparse information systems, other sensor nodes with continuous data streaming to the cloud requires low-power communication techniques, which is one of the foremost requirements in connected bio-sensors due to their energy constraints. an important additional consideration for such nodes is the security of information \cite{DHS_FDA_2019, Spectrum_2019}. With traditional wireless techniques like Bluetooth, an attacker could hack into a pacemaker, insulin pump or brain implant, just by intercepting and analyzing the wireless signals. This hasn't happened in reality yet, but researchers have been demonstrating the risks for over a decade \cite{Das_SciRep_2019, Sen_Spectrum_2020}. Traditional wireless techniques use radiative communication among wearable and implantable devices using electro-magnetic (EM) fields. Owing to such radiative nature of conventional wireless communication, EM signals propagate in all directions, inadvertently allowing an eavesdropper to intercept the information. In this context, the human body, primarily due to its high water content, has emerged as a channel for low to medium-loss transmission, thereby enabling an energy-efficient means for data transfer, termed human body communication (HBC). However, beyond the Electro-Quasistatic (EQS) range of frequencies (frequencies higher then a few 10's of MHz, where the wavelength becomes similar to the dimension of the human body, effectively making the human body a radiative antenna), conventional HBC implementations suffer from significant EM radiation which also compromises security.

\begin{figure}[htbp]
\includegraphics[width=0.88\columnwidth]{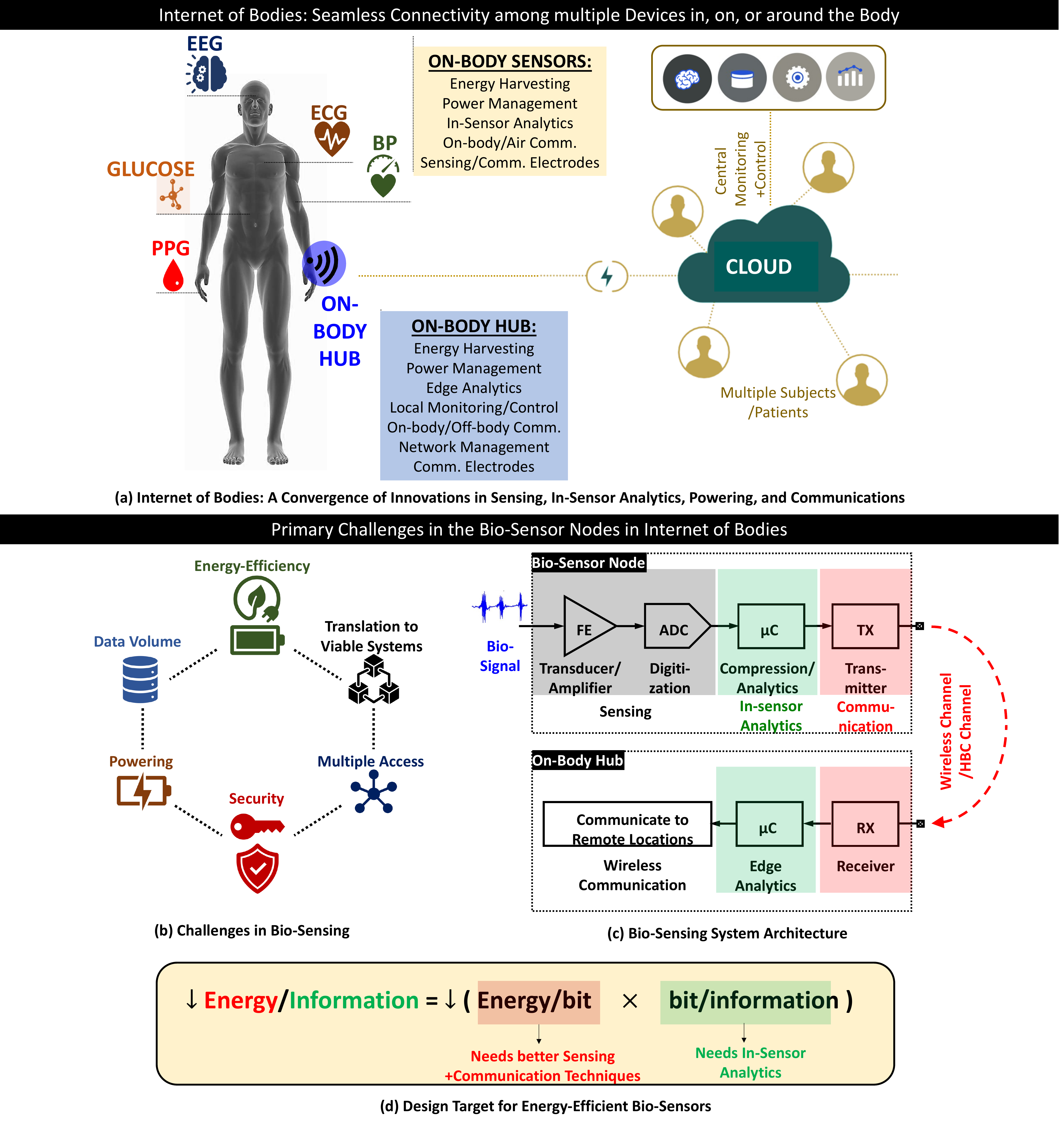}
\caption{\centering (a) The concept of Internet of Bodies (IoB) and its application in bio-sensing; (b) The six challenges in modern bio-sensing; (c) An example of a traditional bio-sensing architecture; (c) Design philosophy for designing energy-efficient bio-sensors. [**Note to Annual Reviews: We created this figure for this article; it is not based on any previously published image.**]}
\label{fig1_motivation}
\end{figure}

Low-power and secure Communication techniques (such as EQS-HBC or MedRadio), along with Spatio-Temporal ISA and Compressive Sensing (CS) has shown immense promise toward building a virtually zero-power, secure network of biomedical sensor nodes for applications including continuous monitoring, brain-machine interfaces and closed-loop bio-electronic medicine. The extremely low power numbers of EQS-HBC \cite{Maity_WRComm_2021, Chatterjee_VLSI_2021, Chatterjee_NatE_2022} can enable perpetual wearable and implantable devices with simultaneous sensing, computation, communication, powering and stimulation for a myriad of bio-sensors.
This ongoing effort toward developing a secure, low-power connectivity solution for multiple sensors on, around, or in the body is being termed as the Internet of Bodies (IoB) \cite{IoB_2020, CIoB_2021}, as shown in \textbf{Figure \ref{fig1_motivation}(a)}. The various sensors on the body, such as Blood-Pressure (BP), Electrocardiogram (ECG), Electroencephalogram (EEG), Photoplethysmogram (PPG) and Glucose sensors collect information about various bio-physical parameters. All of these sensor nodes need to have their individual energy harvesting, power management, in-sensor analytics and on-body communication modalities where they either communicate among themselves, or transmit the collected information to an on-body hub, such as a smartwatch. Again, this on-body hub needs to have its own source of power, processing (called edge analytics), local monitoring and control, network management, and a communication modality to transmit information to the cloud. The cloud performs central monitoring and control, and can send important information and control signals back to the on-body-hub and the sensor nodes. The amount of resource availability (in terms of energy, memory and processing capabilities) keeps on reducing from the cloud to the hub to the individual sensor nodes, and hence the design of these sensor nodes become extremely important for (a) low overall power consumption and (b) high amount of information transfer for unit energy consumption.
\section{CONNECTED BIO-ELECTRONIC SOLUTIONS: Internet of Bodies (IoB)}
\label{Bio_Electronics}

As a subset of the Internet of Things (IoT), IoB represents the network of tiny devices on, around or in the human body comprising of functions such as sensing, analytics, communication, actuation, powering and harvesting. These nodes range from connected health care devices such as continuous glucose monitors (CGM) with connected insulin pumps, connected pacemakers, and ingestible pills to consumer electronics devices such as smartwatches, wireless earpods, AR/VR headsets to name a few. For the rest of the paper, we shall focus on the application of IoB for connected bio-nodes, and discuss the various challenges and opportunities.
\subsection{Challenges in today's Connected Bio-sensors}
Primary challenges in connected bio-sensors are shown in \textbf{Figure \ref{fig1_motivation}(b)}, and are listed below:
\begin{itemize}
\item \textbf{Energy Efficiency:} The wearable bio-sensors (on or around the human body) need to operate with extremely low power consumption for extended battery-life. For implanted devices, the requirement of low energy consumption is even more stringent, as these devices either need to operate with harvested power ($\approx$ 10-100's of $\mu$W), or with a mm$^3$-sized battery which will only hold $\approx$ 2 J of energy, assuming a state-of-the-art energy density of 2 kJ/cm$^3$ \cite{Wiki_Li_2021}. This means that even with an average power consumption of 10 $\mu$W, the battery will run out in $<$3 days, requiring surgery to replace the battery. As a result, power consumption of the implanted bio-node must be less than the limits posed by various modes of energy harvesting.

\item \textbf{Data Volume:} Modern bio-sensors can create huge volume of data during continuous operation. For example, a single-channel neural sensor sampling neuronal activity (up to 10 kHz in frequency) at Nyquist Rate, and with a 16-bit analog to digital converter (ADC), generates data at 10$\times$2$\times$16=320 kbps. With 32 parallel channels, the bit rate increases to 10.24 Mbps, requiring extremely fast data transfer. Assuming that the data transfer happens through traditional radio-frequency (RF) communication (which requires $\approx$ 1 nJ/bit \cite{Sen_DAC_2016}), the communication energy itself will become $>$ 10 mW, which cannot be supported with any form of harvested energy. This means that the data volume needs to be reduced either through compressive sensing \cite{Candes_CS_2005, Donoho_CS_2006}, or through digital compression \cite{Chatterjee_VLSI_2021}, or through some other form of in-sensor analytics.
\item \textbf{Security:} Small form-factor bio-sensors have limited  resources in individual nodes and can hence support only a subset of any intended security feature, such as software encryption. This makes these devices extremely prone to privacy attacks \cite{Jha_IoT_2017}, and needs advanced design methodologies involving building the security features in the hardware itself (i.e. hardware security) \cite{Das_DnT_2021, Das_TCAS_2018, Das_JSSC_2021, Das_Stellar_2019}, and utilizing hardware physical properties for enhancing security \cite{Chatterjee_JIoT_2019, Bari_IMS_2021}.
\item \textbf{Powering:} As explained during the discussion on energy efficiency, powering the wearable/implantable bio-sensors pose significant challenges with the achievable energy consumption from the sensors and today's battery technology, requiring frequent replacement of the batteries. Energy harvesting techniques involving near-infrared (NIR) light \cite{Blaauw_ACS_2021, Blaauw_VLSI_2021}, ultrasonic \cite{UCB_NeuroMethods_2015, UCB_Neuron_2016, UCB_ISSCC_2019, UCB_JSSC_2019, Verhelst_BioCAS_2019, Verhelst_TBioCAS_2019}, thermo-electric \cite{Ren_SciAdv_2021}, RF/Inductive \cite{Shepard_NatBio_2017, Mercier_NatE_2021, Mohseni_BioCAS_2018_1}, magnetoelectric \cite{Yang_ISSCC_2020, Robinson_NER_2021, Robinson_Neuron_2020}, capacitive \cite{Mohseni_ISSCC_2020, Mohseni_JSSC_2019, Mohseni_BioCAS_2018_2} and human body-coupled/electro-quasistatic \cite{JYoo_ISSCC_2020, JYoo_NatE_2021, Modak_CICC_2021, Modak_JSSC_2022, Chatterjee_VLSI_2021, Chatterjee_NatE_2022, Je_TBioCAS_2022}  modes of energy transfer have been explored in the recent years to solve the challenge of powering the bio-sensors. However, the highest possible harvested power that is achieved through these methods is usually $\approx$ 100$\mu$W for reasonable form-factors and channel lengths, garnering significant interest in developing sub-50 $\mu$W bio-nodes in recent years.
\item \textbf{Multiple Access:} Another important consideration in developing a network of bio-nodes is the need for the channel (air for RF, human body for HBC) to be accessed by multiple devices, which may operate at the same time, or using the same frequency band. This may require (a) intelligent methods of frequency allocation when multiple devices access the channel at the same time, or (b) duty cycling with separate communication slot allocation for multiple devices when they access the channel using the same frequency band, or (c) proper use of code-division multiple access when multiple devices access the channel using the same frequency band at the same time.
\item \textbf{Network of Nodes for Multi-Function Operation:}
Multi-functional operation is another important requirement of future bio-sensing and actuation systems. For example, patients with certain neurological disorders can benefit from Neuroprosthesis techniques to restore movement in paralyzed muscles. These systems traditionally focus on a single function with specific motor assistance such as hand grasp. However, paralysis often impacts multiple aspects of life that people want to be restored \cite{Kilgore_TBioCAS_2021}. Implementation of a separate system for each function often becomes impractical since each device requires its own interface as well as real-estate on the body. Additionally, heterogeneity across pathologies makes it difficult to use a single approach that is applicable to everyone. Therefore, a modular, scalable system that can be adapted based on an individual’s unique needs facilitates improving the quality of life.
\item \textbf{Translation of novel architectures to viable systems:} Myriads of new circuits, architectures and techniques have been proposed in the last decade for powering, sensing and communication in the bio-nodes. However, translation of these techniques into viable systems that could be used for long-term monitoring is one of the major considerations that is often overlooked, and requires further research and evaluation to assess their practicality and safety.
\end{itemize}

\textbf{Figure \ref{fig1_motivation}(c)} also shows the traditional bio-sensing architecture, comprising of (a) a bio-sensor node and (b) an on-body hub, with which the bio-sensor node communicates. The bio-sensor node consists a front-end (FE) which amplifies and filters the incoming analog bio-physical signal, followed by an ADC for digitizing the analog signal. A microcontroller ($\mu$C) can be optionally used for in-sensor analytics, which performs some form of compression/analytics on the digitized data to reduce the amount of data volume to be communicated. Finally, a communication transmitter sends the data through a channel (this could be the wireless channel for EM communication, or the human body itself for EQS-HBC), which is received at the on-body hub. In a bio-sensing scenario, usually the bio-sensor node is an extremely small wearable device/implant, and has low amount of resources (energy, memory and computation power). On the other hand, the on-body hub is expected to have a higher amount of resources due to the asymmetric network configuration.

The total amount of energy consumed in the bio-node is given by Eq. \ref{ETot}.
\begin{equation}
    \text{E}_\text{Total} = \text{E}_\text{Sensing} + \text{E}_\text{In-Sensor-Analytics} + \text{E}_\text{Communication}
    \label{ETot}
\end{equation}

where
\begin{equation}
    \text{E}_\text{Sensing} = \text{bit}_\text{sensed} \times \text{E}/\text{bit}_\text{Sensing}
\end{equation}

\begin{equation}
    \text{E}_\text{In-Sensor-Analytics} = f\{\text{bit}_\text{sensed}\}
\end{equation}

and
\begin{equation}
    \text{E}_\text{Communication} = \text{bit}_\text{communicated} \times \text{E}/\text{bit}_\text{Communication}
\end{equation}

For a bio-sensing system, the primary goal is to minimize the total energy, $\text{E}_\text{Total}$ while maximizing the amount of information transmitted. Hence, the design philosophy (\textbf{Figure \ref{fig1_motivation}(d)}) for energy-efficient bio-sensors is usually based on reducing the energy consumption per unit amount of information eventually transmitted, which can be achieved in 2 ways - (a) reduction in energy per bit (requires better sensing and communication techniques, which will be discussed in Sections \ref{sensing_techniques} and \ref{communication}, respectively), (b) reduction in bits per unit amount of information (requires in-sensor analytics that be discussed in Section \ref{ISA}).

\subsection{Societal and Medicinal Impact of Connected bio-sensors: Wearable and Implantable Devices}
The expenditure toward wearable and implantable healthcare devices have become a significant portion of the overall expenses towards healthcare in the U.S. In 2019, U.S. healtcare grew by 4.6\% and reached 3.8T USD, which was 17.7\% of the gross domestic product (GDP) \cite{CMS_NHED_2019}. These expenses are projected to grow to $>$ 5.1T USD by 2023, out of which $>$ 15\% will be spent on wearable and implantable healthcare devices. The continuous monitoring systems are expected to improve at-home patient care and diagnostics, while electroceuticals (including neuromodulation techniques) will provide opportunistic and appropriate neurostimulation which promises to replace/augment pharmaceuticals.

\subsection{Energy Requirements in Bio-sensor Nodes}

The energy requirements in a bio-sensor is often dominated by the energy requirement for communication \cite{Chatterjee_DnT_2019, Cao_VLSI_2020}. Assuming the energy efficiency of wireless communication to be $\approx$ 1 nJ/bit \cite{Sen_DAC_2016}, the transmitter would consume $\approx$ 1 mW power for a nominal data rate of 1 Mbps (which can be expected for a multi-channel neural recorder, for example). Additionally, traditional bio-nodes would also consist an ADC for digitization and an optional compressive sensing unit for data volume reduction. Both of these modules consume $\approx$ 50-100 $\mu$W power each at the target data rates for state-of-the-art implementations \cite{Chatterjee_VLSI_2021, SAR_ADC_2019}. Hence, the overall power consumption in the transmitter would be $\approx$ 1.2 mW. This is $>$ 3 $\times$ higher than the known techniques of energy harvesting for mm to cm-scale implants. Consequently, reducing the communication power has gained significant focus in the recent years. A $<$ 50 pJ/bit (at 1 Mbps) wireless communication technique involving both MedRadio and Human Body Communication (HBC) standards was shown in \cite{Chatterjee_RFIC_2020}, making the communication power similar to that of digitization and compression. 
 
\subsection{Achieving Ultra-Low-Power Bio-Sensor Nodes}
To achieve ultra-low-power bio-sensing, all the submodules in the bio-node (sensing, computation/analytics and communication) need to contain certain innovations in terms of their energy efficiencies. Furthermore, powering and network-level interaction techniques need to be implemented so that the bio-node can operate continuously without manual intervention.  
\subsubsection{Low-Power Sensing Techniques}
Low-power sensing techniques including compressive sensing, time-domain sensing and collaborative sensing has gained significant interest in the recent years. Compressive Sensing utilizes the inherent sparsity of bio-physical signals and reduces the amount of digitized communication payload, thereby reducing communication energy. Time-domain sensing can achieve moderate resolution (12-18 bits) for bio-sensors, without using complex voltage-mode or current-mode architectures. This is possible because most of the bio-physical signals are of low frequency (Hz-kHz range) and hence the availability of `time' can be utilized to average the noise and achieve high signal to noise ratio (SNR), which eventually results in high resolution. The methods related to compressive sensing and time-domain sensing are explained in detail in Section \ref{sensing_techniques}.

\subsubsection{In-Sensor Analytics for reducing Communication Power}
In-sensor analytics (ISA) can also reduce the communication energy, by either making the communication event-driven, or by compressing the sensed data \cite{Chatterjee_JIoT_2021}. Event-driven communication enables non-uniform duty-cycling during data transmission, which reduces the overall communication energy without losing important information. Some of the ISA techniques applicable for bio-sensors are explained in detail in Section \ref{ISA}.

\subsubsection{Communication and Powering}
 In most cases, traditional radio-frequency (RF) communication around the body consumes $>$ 1 nJ/bit, which poses serious constraints on high-speed communication ($>$ 1 Mbps or more, which requires $>$ 1 mW power), as energy harvesting techniques can only provide $\approx$100 $\mu$W power in most cases. The challenge is even more for implants as replacement of batteries require surgery and hence energy harvesting becomes the only viable option \cite{Singer_Health_2021}. Recent techniques in low-power communication are explained in detail in Section \ref{communication}, while the powering techniques are shown in section \ref{powering}.

Previous works from several research groups in all of these different domains have been considered in this paper, and have been summarized in Table \ref{fig2_Table}.

\begin{table}[htbp]
\caption{\centering Popular Research Directions of the constituent elements of bio-sensing nodes}
\vspace{-5mm}
\footnotesize
\begin{center}
\begin{tabular}{@{}l|c|c|c@{}}

\hline
\hfil &\hfil Modality/Architecture&\hfil References&\hfil Salent Features\\\hline \hline

\centering &1. High-Impedance, Chopper&\cite{Yazicioglu_2007, Enz_1996, Denison_2007}& Low-Frequency\\
\centering &Stabilized Front-ends&& Low-Noise.\\
\centering &2. Voltage vs. Time-&\cite{Chatterjee_CICC_2019, Chatterjee_JSSC_2020}& Low-Frequency\\
\centering Sensing&Domain Architectures&& Low-Power.\\
\centering &3. Compressive Sensing&\cite{CS_MRI_2007, AA_TCASII_2017, NV_MMADC_2015, AA_BioCAS_2017}& Sparse Signals,\\
\centering &&\cite{Gaurav_JSSC_2022, Verhelst_ESSIRC_2018, Trakimas_TCAS_2011, Martins_TVLSI_2017}&Low-Power.\\
\centering &4. ADC-less Sensing&\cite{JoshuaSmith_USENIX_2018, Chatterjee_ESSCIRC_2022}& Low-Frequency,\\
\centering &&&Extremely Low-Power.\\
\hline

\centering &1. MedRadio/Low-&\cite{Chatterjee_RFIC_2020, Mondal_RFIC_2020, Lee_JSSC_2019}& Low-Interference,\\
\centering &Power Wireless&&Low-Power.\\
\centering &2. EQS Capacitive HBC&\cite{Maity_BodyWire_2019, Lucev_HBC_2012, Bae_HBC_2012, Park_TBME_2017}& Ultra Low-Power,\\
\centering &&\cite{Maity_TBME_2019, Nath_TCAS2_2020, Maity_EMBC_2018, Chatterjee_ISSCC_2022}&Physically Secure.\\
\centering Communication&3. EQS Galvanic HBC&\cite{Chatterjee_VLSI_2021, Chatterjee_ESSCIRC_2022, Modak_TBME_2021}& Short Distance\\
\centering &\& Bi-Phasic HBC&&(few cm).\\
\centering &4. MQS (Magnetic) HBC&\cite{Mag_EMBC_2015, Mag_ISSCC_2019, Mag_TBME_2021}& $\approx$ Near Field, No\\
\centering &&&effect of Body.\\
\centering &5. EM-wave HBC&\cite{Tochou_JSSC_2022}& High DR, Good Energy\\
\centering &&&Efficiency, Not as Secure.\\
\hline

\centering &1. Event-driven&\cite{NV_MMADC_2015, AA_BioCAS_2017}& Asynchronous,\\
\centering &Compression&&Low-Power.\\
\centering In-Sensor Analytics&2. Spike and/or&\cite{Sodagar_JSSC_2009, Sodagar_TBioCAS_2014, Mich_ISSCC_2020, Schwartz_ARNeuro_2004}& Asynchronous,\\
\centering &Anomaly Detection&\cite{Schwartz_TNSRE_2003, Wise_JSSC_2005, Harrison_JSSC_2007}&Ultra Low-Power.\\
\centering &3. Learning-based&\cite{Mohseni_Asilomar_2020, Mohseni_EMBC_2021}& Low-Power\\
\centering &Analytics&&Machine Learning.\\
\hline

\centering &1. Near-Infrared&\cite{Blaauw_ACS_2021, Blaauw_VLSI_2021}& Lowest Form-factor,\\
\centering &&&Needs Repeater.\\
\centering &2. Ultrasound&\cite{UCB_NeuroMethods_2015, UCB_Neuron_2016, UCB_ISSCC_2019}& Low Form-factor at Low\\
\centering &&\cite{UCB_JSSC_2019, Verhelst_BioCAS_2019, Verhelst_TBioCAS_2019}&Freq., Needs Repeater.\\
\centering &3. RF \& Inductive&\cite{Shepard_NatBio_2017, Mercier_NatE_2021, Mohseni_BioCAS_2018_1}& Large Coil/Antenna,\\
\centering Wireless Powering&&&Traditional Architecture.\\
\centering &4. Magneto-Electric&\cite{Yang_ISSCC_2020, Robinson_NER_2021, Robinson_Neuron_2020}& Better Power Transfer\\
\centering &&&Efficiency than RF, Safe.\\
\centering &5. Capacitive&\cite{Mohseni_ISSCC_2020, Mohseni_JSSC_2019, Mohseni_BioCAS_2018_2}& Larger Form-factors,\\
\centering &&&Small Distances.\\
\centering &6. EQS&\cite{JYoo_ISSCC_2020, JYoo_NatE_2021}&  Full-Body Powering\\
\centering &&\cite{Modak_CICC_2021, Chatterjee_VLSI_2021}&for Wearables.\\
\hline

\end{tabular}
\end{center}
\label{fig2_Table}
\normalsize
\end{table}
\vspace{-10mm}
\section{ULTRA-LOW-POWER SENSING: Towards Batteryless, Perpetual Operation}
\label{sensing_techniques}

In order to minimize the Energy consumption per unit information, energy-efficient techniques for sensing and communication needs to be explored, along with data volume reduction using ISA. This section talks about recent methods in energy efficient sensing.

\begin{figure}[htbp]
\includegraphics[width=\columnwidth]{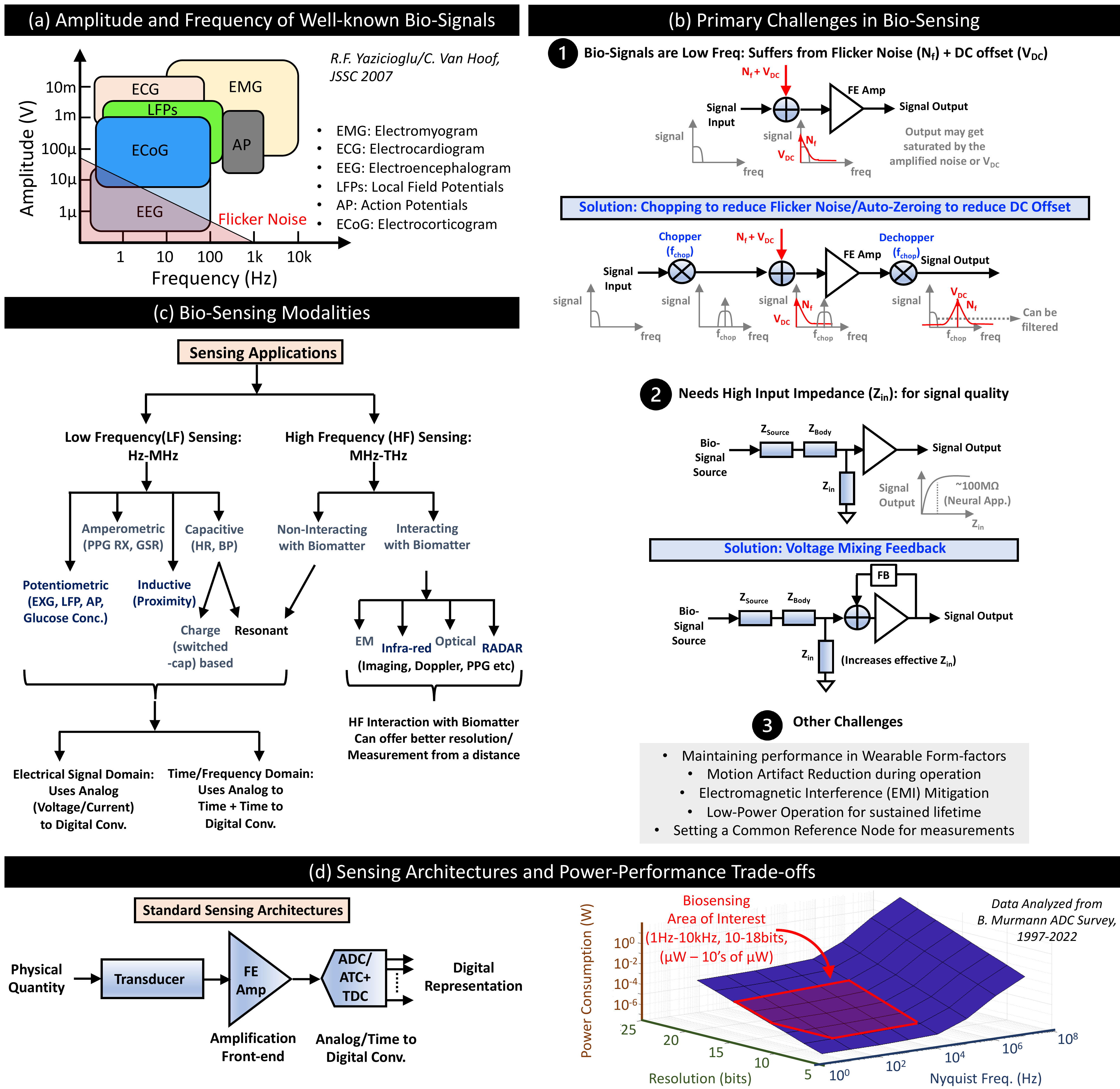}
\caption{\centering (a) Amplitude and Frequencies of well known bio-signals are shown, indicating the requirement of low frequency (Hz-10kHz) sensing of low-amplitude (up to $\mu$V) signals. (b) The primary challenges in such systems arise from low-frequency Flicker Noise that coincides with the frequency of the bio-signals, and requirements of high input impedance and high resolution. However, the power consumption of (c) different modalities or (d) architectures usually lie within a few 10's of $\mu$W range. [**Note to Annual Reviews: We created this figure for this article; it is not based on any previously published image.**]}
\label{fig3_Sensing}
\end{figure}

Many of the naturally occurring bio-physical signals are slowly-varying, such as ECG, EMG, EEG, and ECoG. Most of the energy content in these signals are contained within low frequencies (1Hz-10kHz range), as shown in \textbf{Figure \ref{fig3_Sensing} (a)} \cite{Yazicioglu_2007}. However, the resolution as well as dynamic range requirements for these applications can be large (12-16 bits for traditional bio-sensing applications, and up to 20 bits for extremely low-amplitude EEG). Traditional voltage-mode and current-mode ADC designs in such high-resolution applications become severely limited by (a) the low-frequency flicker noise originated from the amplifiers and analog to digital converters (ADC), as well as by (b) the requirements of high input impedance, (c) low DC offsets, and (d) low power consumption, as presented in \textbf{Figure \ref{fig3_Sensing} (b)}. Techniques such as chopping helps mitigating the challenges due to flicker noise. The flicker (or 1/f) noise arise from the Si-SiO$_2$ interface non-idealities in traditional CMOS-based circuits (used to amplify and digitize the bio-signals), and is most prevalent at lower frequencies. Chopping up-converts the input bio-signals to higher frequencies, so that the low-frequency flicker noise at the input of the CMOS amplifiers do not affect the bio-signals anymore \cite{Enz_1996, Denison_2007}. Similarly, auto-zeroing techniques or DC servo-loops \cite{Enz_1996, Denison_2007, Markovic_2017} help reducing and DC offsets resulting from the electrodes used for acquisition of the bio-signals. Voltage mixing/positive feedback techniques help improving the input impedance of the sensor \cite{Denison_2007, Markovic_2017}, so that most of the signal is available at the input of the sensor.

The sensing itself can be divided into (\textbf{Figure \ref{fig3_Sensing} (c)}) low-frequency (LF: Hz-MHz) and high-frequency (HF: MHz-THz) techniques. LF techniques can be Potentiometric, Amperometric, Inductive, or Capacitive, each of which could be read out either in the electrical signal domain (voltage/current), or using time-domain techniques (frequency/time period). The HF techniques can either be non-interacting with the biomatter (such as resonant techniques to read out a capacitive sensor), or interacting with the biomatter (such as imaging, doppler, or PPG systems) that offer better resolution, and sometimes, measurement from a distance \cite{Mehrotra_Sensors_2019}. If we plot the power consumption of such sensing front-ends with respect to the ADC resolution and Nyquist frequency \cite{Murmann_2022}, we shall find that for the bio-sensing area of interest, the power consumption is usually limited in the range of a few $\mu$W to a few 10's of $\mu$W, which is significantly lower than the commonly observed mW power consumption of communication. 

In recent years, time-based ADCs have gained significant interest, because of their ability to utilize the availability of time (since signals are of very low frequency) in an energy-resolution scalable manner as shown in \cite{Chatterjee_DnT_2019, Chatterjee_CICC_2019, Chatterjee_JSSC_2020}. For high resolution requirements, the signal to be sensed is converted into an equivalent frequency, and is simply observed using a counter for longer amount of time for a change in the average frequency. For low resolution requirements, the frequency is observed for a shorter amount of and then can be turned-off (through duty cycling) for saving energy.

Even though time-based methods ensure energy-resolution scalability within a certain range, the resolution cannot be made infinitely high by measuring (or averaging the noise) for longer time. The ambient noise statistics, PVT variation and jitter accumulation in the ring oscillator would limit the achievable resolution, out of which jitter accumulation is shown to be the dominant factor in \cite{Chatterjee_CICC_2019, Chatterjee_JSSC_2020} in a controlled environment. The scaled quantization error (SQE) in measuring a fixed frequency within a pre-defined amount of time goes down with the time of measurement. However, the accumulated jitter from the ring oscillator goes up with the total time of measurement. If the slope of the linear plot of accumulated jitter vs measurement time is $k$, then the achievable resolution is shown to be limited to $\log_2{(1/k)}$ bits.

\subsection{Emerging Techniques in Ultra-Low-Power Sensing}
\subsubsection{Compressive Sensing}
Compressed-domain sensing/compressive sensing (CS) \cite{Candes_CS_2005, Donoho_CS_2006} is a mathematical tool in signal processing that defies the Shannon-Nyquist sampling theorem by sampling a sparse signal at a rate lower than the Nyquist rate, while still being able to reconstruct the signal with negligible error. Since its inception, CS has found multiple applications including medical imaging \cite{CS_MRI_2007}, in-sensor analytics \cite{AA_TCASII_2017}, healthcare \cite{NV_MMADC_2015, AA_BioCAS_2017}, neural applications \cite{Chatterjee_VLSI_2021} and audio acquisition \cite{Gaurav_CICC_2021, Gaurav_JSSC_2022} . CS algorithms assume that the signal to be sampled has a sparse representation, and it was shown that sparse signals with randomly (from an i.i.d. Gaussian distribution) undersampled data can be recovered with low error by formulating it as an optimization problem. Hence, the advantage of compressive sensing is two-fold: (a) CS allows a lower sampling rate which reduces the power consumption in the analog to digital converter (ADC) and clock generation circuitry, (b) compression creates smaller amount of data with rich information-content which reduces the burden on the subsequent processing and communication modules. Since many of the bio-physical signals can be represented in sparse form \cite{CS_Survey_2013}, it is possible to leverage the superior energy efficiency of CS for bio-sensing. Two comprehensive reviews on CS can be found in \cite{CS_Survey_2013}  and \cite{CS_Survey_2018}. In \cite{Verhelst_ESSIRC_2018}, a sub-$\mu$W compressive sensing hardware is presented in 65nm CMOS technology with online self-adaptivity for incoming signals with varying sparsity. Initial efforts on self-adaptivity was earlier demonstrated in \cite{Trakimas_TCAS_2011} using an asynchronous ADC with adjustable sampling rate, and in \cite{Martins_TVLSI_2017} using temporal decimation and wavelet shrinkage. Both of these techniques were utilized with specific incoming signals. \cite{Verhelst_ESSIRC_2018}, on the other hand, offers a more general technique that exploits the online sensory data statistics for dynamic reconfiguration (in terms of compression algorithm, compression harshness and sampling frequency). Recent works \cite{Chatterjee_VLSI_2021, Gaurav_CICC_2021, Gaurav_JSSC_2022} demonstrate a fully digital CS subsystem, equipped with an on-chip 2-stage sparsifier and dual varying-seed-PRBS sensing-matrix generator, with variable compression factor (CF) from 5X to 33.33X. The 2-stage Discrete Wavelet Transform (DWT) based sparsifier ensures that the CS module works effectively for both sparse and non-sparse signals.

\subsubsection{ADC-less Sensing}

Traditional sensor nodes consist a transducer, analog amplifiers, an ADC, an optional digital compression unit and a digital communication module. The ADC consumes a large amount of power, while creating multiple bits from each sample of the input data, which increases the burden on the communication module. The digital compression unit reduces this communication burden by lowering the number of bits to be transmitted, but consumes additional energy. Instead, an ADC-less architecture can be used, which creates a pulse-width modulated (PWM) signal corresponding to each input sample. This encodes the analog information into the analog pulse-width of the signal, which can still be communicated using a digital-friendly transmitter, since the amplitude of the PWM signal remains rail-to-rail. This method was proposed in \cite{JoshuaSmith_USENIX_2018}, while the first IC-implementation for resource-constrained wireless neural implants has recently been shown in \cite{Chatterjee_ESSCIRC_2022}, obviating the need for the power-hungry ADC and digital compression modules.

Realizing that all bio-sensing applications involve asymmetric resource distribution (the sensor nodes/transmitters usually have small amount of resources, while the hub/receiver can have more amount of resources in terms of energy, memory and processing capabilities), reducing the energy consumption at the node transmitter using such ADC-less architectures would be extremely useful, and are one of the directions of future research. 

\subsubsection{Collaborative Sensing}
Collaborative Wireless Sensor Networks (CWSN) \cite{CollabSense_2011, CollabSense_2018} can sense an analog signal using multiple sensors, utilizing collaborative efforts among the sensor nodes and their communication with each other/with the cloud.  The trade-offs and power optimization in such networks have been analyzed in \cite{Chatterjee_JIoT_2021} for large-area IoT test-beds, and can be utilized for bio-sensing applications as well.

The above techniques help in reducing the energy-cost of sensing. However, the energy consumption during communication is usually the dominant component in a bio-sensor node, which is why it is extremely important to chose a data transmission modality that helps in reducing the overall system-level power consumption.
\section{ULTRA-LOW-POWER COMMUNICATION: Designing Perpetual Systems}
\label{communication}

\subsection{The journey from Personal Area Networks to Body Area Networks to IoB}

\begin{figure}[htbp]
\includegraphics[width=\columnwidth]{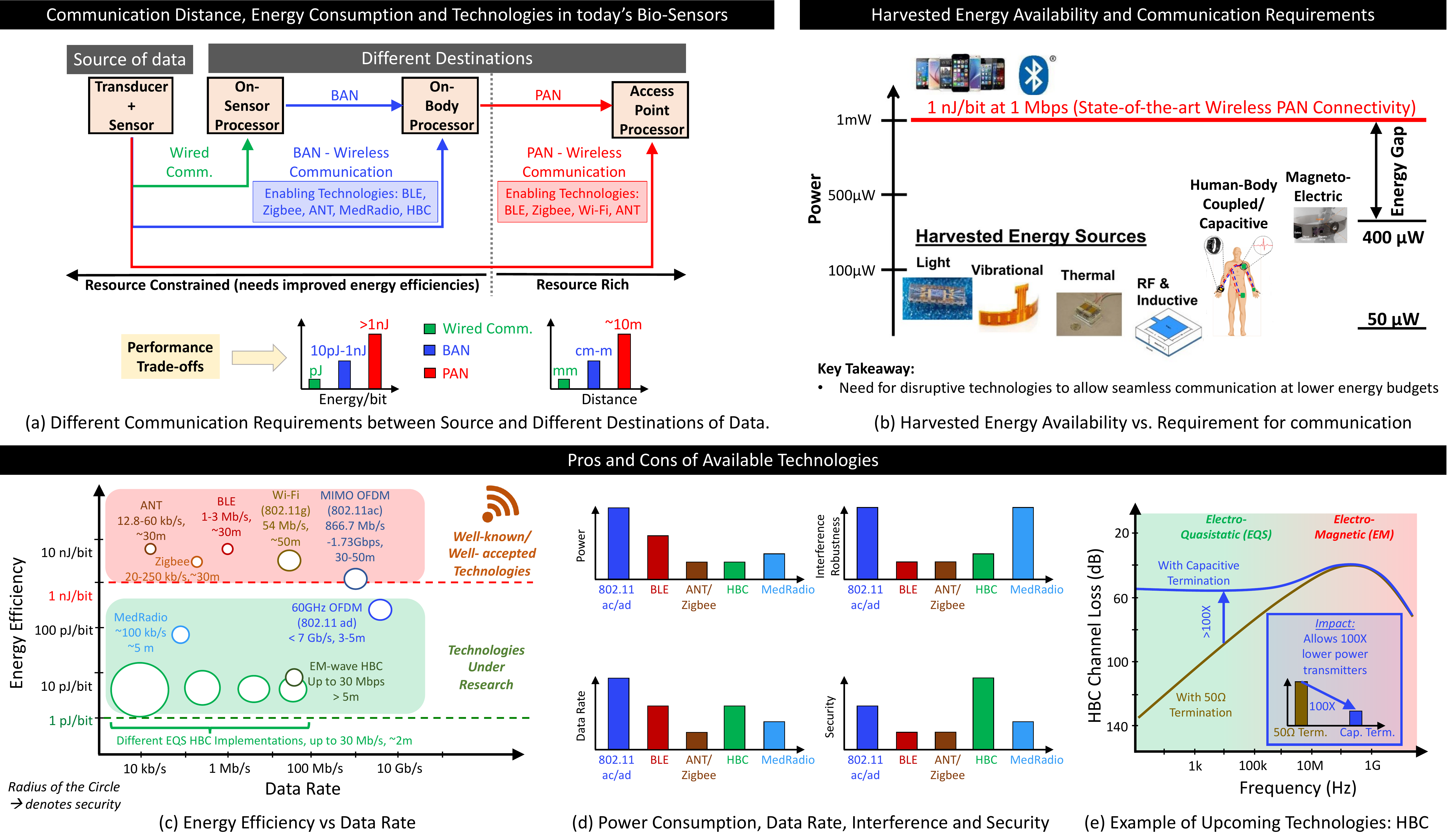}
\caption{\centering (a) Communication requirements and technologies used between the source and different destinations of data in IoB; (b) The gap between harvested energy numbers and traditional wireless communication requirements \cite{Chang_JETCAS_2018}; (c) Energy-efficiency vs. data rate of available technologies, (d) Comparing available technologies w.r.t. their power consumption, data rate, interference rejection capabilities and security; (e) Benefits of upcoming capacitively terminated HBC technology. [**Note to Annual Reviews: part (b) is adapted from \cite{Chang_JETCAS_2018} for which Chatterjee B and Sen S are co-authors; rest of the figure is not based on any previously published image.**]}
\label{fig_Comm}
\end{figure}

Communication in IoB can be categorized w.r.t. the source of data (which is the bio-sensing node) and the different destinations, as shown in \textbf{Figure \ref{fig_Comm}(a)}. The destination could be 1) an on-sensor processor: this could be at a distance of a few mm-cm, and the communication would only require pJ/bit energy, since wired communication can be used in this scenario; 2) an on-body processor/aggregator: this could be a smart watch, for example, at a distance of a few cm-m, and would require body-area network (BAN) based communication, for which the available technologies include bluetooth low-energy (BLE), Zigbee, ANT, MedRadio, HBC etc., which could consume 10-1000 pJ/bit energy depending on the design; 3) a remote processor at an off-body access point: this could be at a distance of a few m, and would require a personal-area network (PAN) for communication, for which the available technologies include bluetooth low-energy (BLE), Zigbee, Wi-Fi, ANT, and so on, all of which could consume $>$1 nJ/bit energy.

Most of the wireless communication techniques employed for for BAN and PAN (for example, BLE, Zigbee, ANT or Wi-Fi) typically consume 2-3 orders of magnitude more power than sensing and computation, making it the biggest bottleneck in achieving energy-harvested/battery-less sensor nodes \cite{Chatterjee_DnT_2019, Cao_VLSI_2020, Chang_JETCAS_2018, Cao_JSSC_2022}. For an Mbps communication link, this results in a power consumption of a mW or more, as shown in \textbf{Figure \ref{fig_Comm}(b)}. However, it must be kept in mind that eventually all bio-sensors are envisioned to perpetually operate using harvested energies from various sources (light/ vibrational/ thermal/ RF/ inductive/ magnetic/ human-body-coupled), which are limited to a few 100 $\mu$W even in the most favourable conditions. This means either we need to reduce the communication burden by better ISA techniques, or we need to decrease the pJ/bit of communication itself. One thing that needs to be considered here is that when most of the early research started on low-power BAN in the late 2000's, there was not much distinction between techniques use for BAN and PAN. However, with the widespread use of new/emerging techniques such as extremely low-power MedRadio or HBC, the energy efficiencies in BAN communication has been reduced to $\approx$ 10pJ/bit \cite{Maity_BodyWire_2019, Chatterjee_RFIC_2020}, leading to $\approx$100X power benefits as compared to traditional wireless, bridging the gap between sensing/computation power and communication power. As a result, the design of extremely low-power bio-sensor nodes in future will become a co-design problem among sensing, computing and communication (and perhaps even security), instead of being limited only by the communication power.

The energy-efficiency vs. data rates of various communication technologies are plotted in \textbf{Figure \ref{fig_Comm}(c)}, with the radius of the circle indicating the qualitative security of the communication scheme. Almost all well-known/traditional wireless techniques consume more than 1nJ/b, and offer low electromagnetic (EM) security, as the EM waves are present even at a distance of 5-10 m, making the signals susceptible to hacking \cite{Das_SciRep_2019, Sen_Spectrum_2020}. More emerging techniques such as low-power MedRadio \cite{Chatterjee_RFIC_2020, Mondal_RFIC_2020}, electro-quasistatic HBC (EQS-HBC) \cite{Maity_BodyWire_2019, Lucev_HBC_2012, Bae_HBC_2012, Park_TBME_2017, Maity_TBME_2019, Nath_TCAS2_2020, Maity_EMBC_2018, Chatterjee_ISSCC_2022, Modak_TBME_2021, Mag_EMBC_2015, Mag_ISSCC_2019, Mag_TBME_2021, Chatterjee_VLSI_2021, Chatterjee_NatE_2022, Chatterjee_ESSCIRC_2022} and EM-wave HBC \cite{Tochou_JSSC_2022} offer improved energy efficiencies because of their 1) lower carrier frequencies, 2) lower output power (MedRadio/EM-wave HBC) and non-50$\Omega$, high-impedance terminations (EQS/EM-wave HBC). Interestingly, EQS-HBC signals stay relatively constrained within the human body, offering better security than EM. The amount of leakage increases with frequency though, which means low-frequency EQS-HBC offer very high physical security. However, since the human-body is susceptible to various environmental interference, the robustness of operation is much worse than some of the standardized techniques such as MedRadio or MIMO OFDM (IEEE 802.11 ac/ad) which were designed for low interference. In \textbf{Figure \ref{fig_Comm}(d)}, we compare and contrast different communication techniques for bio-sensors. As an example modality, \textbf{Figure \ref{fig_Comm}(d)} shows the benefits of EQS-HBC with capacitive termination as an alternative to radio wave-based Wireless Body Area network (WBAN), terminated with a 50-$\Omega$ antenna. In EQS communication, low frequency ($<$1 MHz) electrical signals are communicated between two sensor nodes (transmitter and receiver) placed on the body by utilizing human body as a transmission channel. Capacitive termination at both the TX and RX makes the channel loss a function of a capacitive division, as shown in \cite{Maity_BodyWire_2019}, which means that we have a flatband channel, even at low frequencies, instead of a high-pass channel, which is common in 50-$\Omega$ body-communication systems. The use of lower frequencies for operation, along with a wide bandwidth leads to such improved energy efficiencies for EQS-HBC. EQS communication also shows better performance in terms of physical security as most of the signal is confined within the body. Capacitive \cite{Maity_BodyWire_2019, Lucev_HBC_2012, Bae_HBC_2012, Park_TBME_2017, Maity_TBME_2019, Nath_TCAS2_2020, Maity_EMBC_2018, Chatterjee_ISSCC_2022}, Galvanic \cite{Modak_TBME_2021}, Magnetic \cite{Mag_EMBC_2015, Mag_ISSCC_2019, Mag_TBME_2021} and Bi-Phasic \cite{Chatterjee_VLSI_2021, Chatterjee_ESSCIRC_2022} are the major quasistatic modalities in terms of excitation and termination at the RX and TX.

\subsection{Low-Power Techniques in Wireless Communication for IoB}
\subsubsection{Low-Rate Wireless PAN (LR-WPAN) and Body Area Network (BAN)}
LR-WPAN is a technical standard that is maintained by the IEEE 802.15.4 working group (started in 2003), and focuses on low-power wireless communication techniques using physical layer, data-link layer and network layer optimizations. On the other hand, BAN evolved as part of the IEEE 802.15.6 standard (2011), which aims to provide the low-power features of 802.15.4, while optimizing for short range communication around the body. 
\subsubsection{MedRadio}
Medical Device Radiocommunications Service (MedRadio), previously called the Medical Implant Communication Service (MICS), aims to support proliferation of low-cost medical sensor technologies for diagnostic, therapeutic and monitoring purposes, by dedicating certain low-interference frequency bands, primarily near 400 MHz, although 2.36-2.39 GHz MBAN devices also fall into the MedRadio Category. These standards have been developed during late 1990s-early 2010's, when it became part of IEEE 802.15.6. However, with recent trends in low-power wireless IC design due to digital-friendly high-speed low-power processes, there has been renewed interest in the MedRadio space \cite{Lee_JSSC_2019, Chatterjee_RFIC_2020, Mondal_RFIC_2020}. 
\subsubsection{Human-Body Communication (HBC)}
Human-Body Communication utilizes the conductive properties of the human tissue for communicating signals for devices in, on or around the human body. The use of low-frequency EQS signaling helps in restricting the signals to the body, without a significant amount of leakage, while simultaneously achieving low power consumption due to the use of lower carrier frequencies. In the next subsection, we shall discuss some of the emerging techniques in low-power/energy-efficient HBC.

\subsection{Emerging Techniques in Energy-Efficient HBC}
\subsubsection{Capacitively terminated, Wideband, Low-Frequency EQS-HBC}
Using the human body  as a low-loss broadband communication medium \cite{Park_TBME_2017, Maity_TBME_2019}, $<$ 10 pJ/bit energy-efficiencies similar to wireline IO \cite{IO_ISSCC_2014, Hsueh_ISSCC_2014} can be achieved, while simultaneously achieving high physical security \cite{Das_SciRep_2019}. A sub-10 pJ/b Capacitive HBC link was first shown in \cite{Maity_BodyWire_2019}, with voltage-mode signaling and high-impedance termination that allows broadband communication. The low channel loss and absence of up/down-conversion results in extreme energy efficiencies. The key challenge in broadband HBC comes from the antenna effect in human body which picks up unwanted interference that corrupt the signal. An interference detection and rejection loop using an adaptive notch at the integrating Rx enabled a 6.3 pJ/b transceiver for 30 Mbps data transfer through the body (the energy efficiency was $\approx 100 \times$ lower than traditional WBAN) with an interference robust (can tolerate -30 dB signal to interference ratio) HBC transceiver. \cite{Chatterjee_RFIC_2020} and \cite{Mag_ISSCC_2019} have later shown $<$ 10 pJ/bit transceivers with on-chip clocking architectures. For ultra-low power physiological monitoring and secure authentication, \cite{Maity_WRComm_2021} has recently demonstrated a 415 nW capacitive EQS-HBC link with 1-10 kbps data rates, and exhibits physical and mathematical security through integration of an AES-256 encryption engine along with the HBC transceiver. \cite{Chatterjee_ISSCC_2022} analyzes the theoretical limits of power consumption in such systems and demonstrates adiabatic techniques with a hardware-aware modulation scheme to improve energy efficiencies. Such low power levels can potentially enable the design of battery-less wearable patches in future.

\subsubsection{Multi-mode Resonant HBC for Communication and Powering}
One of the major sources of power consumption in voltage-mode capacitive HBC TX is the inter-electrode parasitic capacitance. \cite{JYoo_ISSCC_2020, JYoo_NatE_2021, Modak_CICC_2021} have recently showed that by using an inductance across this capacitor presents a parallel resonance at the load of the TX, which in turn reduces the power consumption during data communication. Similarly, using a series resonance at the TX output (with a series inductor) increases the output voltage by a factor of Q (where Q is the quality factor of the inductor). Since the power transfer in a voltage-mode capacitive HBC link increases quadratically with the output voltage of the power TX, this method increases the power transferred through the body by a factor of Q$^2$. As compared to \cite{JYoo_ISSCC_2020} and \cite{JYoo_NatE_2021} where the amount of power transferred reduced with the distance between the TX and RX, \cite{Modak_CICC_2021} achieves an almost constant power transfer and efficiency throughout the body as the operation is at a much lower frequency (1 MHz in \cite{Modak_CICC_2021} vs 40 MHz in \cite{JYoo_ISSCC_2020, JYoo_NatE_2021}).
\subsubsection{Bi-Phasic QBC for Powering and Communicating with Deep Brain Implants}
For implantable scenarios, capacitive HBC is not very effective because of an absence of a strong return path from within the body, whereas galvanic HBC consumes higher power as a significant amount of DC current flows into the surrounding tissue if the signal is not DC-balanced. Bi-Phasic QBC \cite{Chatterjee_VLSI_2021} is recently shown to reduce the power consumption of the galvanic modality by using AC coupling at the output. As compared to Galvanic HBC, Bi-Phasic QBC achieves $\approx$ 41$\times$ lower power at 1 MHz EQS frequency.

Even with such low-power, emerging data communication techniques, the energy cost of communication still dominates in a bio-sensor node. Consequently, it is imperative to look at available in-sensor computation techniques, that can reduce the computation burden in the node.
\section{IN-SENSOR DATA ANALYTICS: Computation vs. Communication}
\label{ISA}

As the number of distributed sensors in IoB increases, the total amount of data transfer to the back-end hub/cloud servers are becoming prohibitively large, resulting in network congestion and high energy consumption during data transmission at the sensor node \cite{Cao_IMS_2017}. This motivates the need for in-sensor data analytics which would perform context aware data acquisition with some amount of computation, followed by transmit if necessary, resulting in interesting trade-offs as shown in \textbf{Fig. \ref{comm_en}(a)}. Depending on the amount of in-sensor analytics (ISA) performed in the bio-sensor, the total energy could be dominated by either the communication energy, or the computation energy.

\subsection{Trade-Offs between Computation and Communication}
We can define the computation and communication energies ($\text{E}_\text{computation}$ and $\text{E}_\text{communication}$, respectively) in a bio-sensor node as given in \ref{ISA_eqn}:

\begin{eqnarray}
	\text{E}_\text{computation} = \left(\text{E}_\text{comp}/\text{bit}\right) \times \text{No. of bits switched}\\
	\text{E}_\text{communication} = \left(\text{E}_\text{comm}/\text{bit}\right) \times \text{No. of bits transmitted}
\label{ISA_eqn}
\end{eqnarray}

For digital computation units (which is the conventional implementation), the low-frequency region of $\left(\text{E}_\text{comp}/\text{bit}\right)$ is usually dominated by the static leakage current in the devices used for computation, while the high-frequency region is dominated by the dynamic energy of bit-switching, as demonstrated in \cite{Chatterjee_TVLSI_2019, Chatterjee_ASPDAC_2021}. For optimum energy efficiency, the designer looks for a region where the sum of the leakage and dynamic energies can be minimized. This leads to $\left(\text{E}_\text{comp}/\text{bit}\right)$ being in the range of a few fJ-pJ. On the other hand, as shown in \cite{Chatterjee_DnT_2019, Chatterjee_JIoT_2021}, $\left(\text{E}_\text{comm}/\text{bit}\right)$ is usually determined by the receiver's bit-error rate (BER) sensitivity for a particular Data Rate, the communication channel loss, and the transmitter's efficiency, which usually results in $\left(\text{E}_\text{comm}/\text{bit}\right)$ being in the 100s of pJ to 1 few nJ range for modern implementations. 

\begin{figure}[htbp]
\includegraphics[width=0.95\columnwidth]{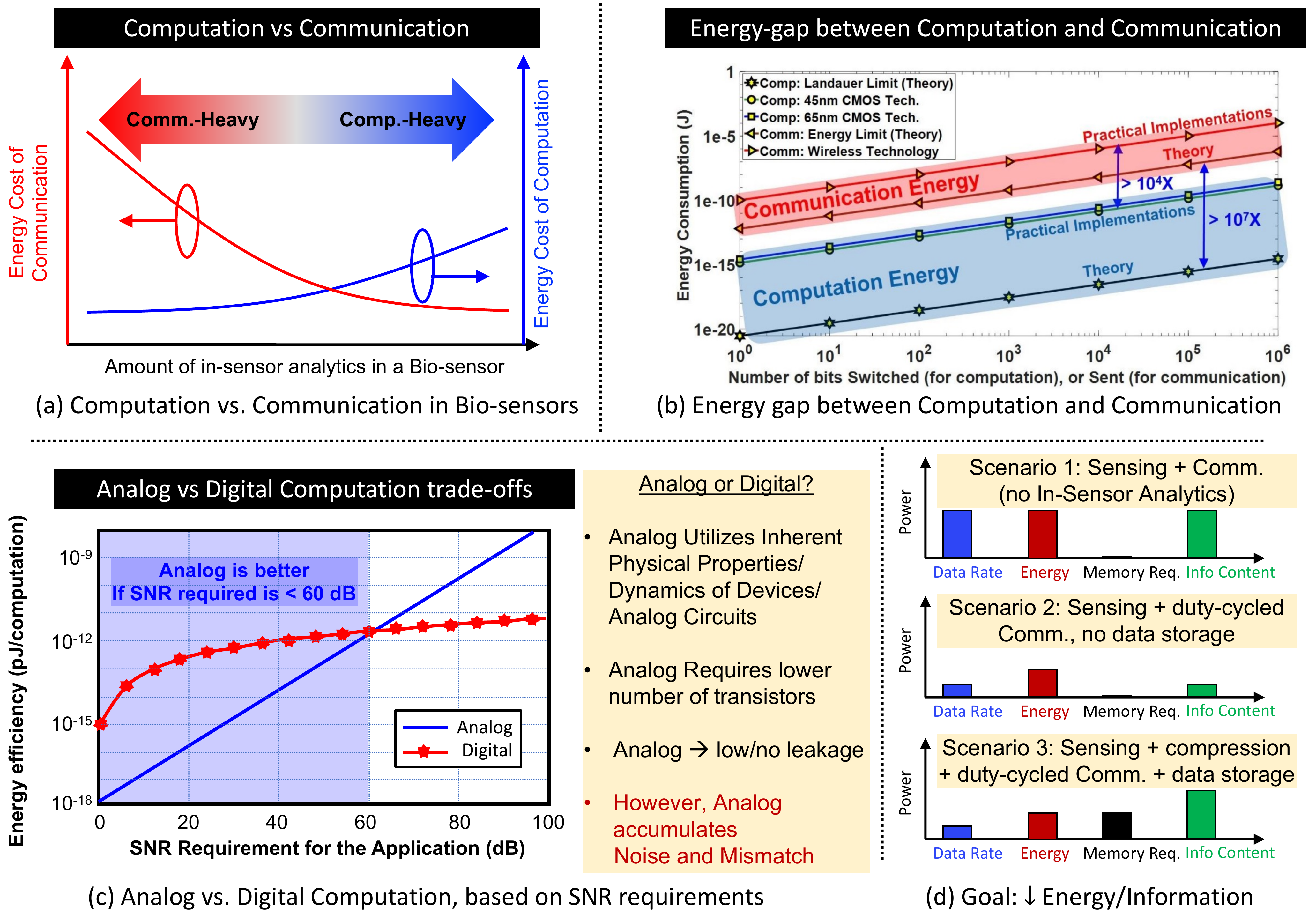}
\caption{\centering (a) Computation vs. Communication energies in a bio-sensor, based on amount of in-sensor analytics performed. (b) Comparison of Communication and Computation energies (both theoretical and from standard implementations \cite{Chatterjee_DnT_2019}) that show that Communication energy is $\approx 10^4$ times more than Computation energy ( with same number of bits). (c) Analog vs. Digital Computation trade-offs \cite{Sarpeshkar_AD_1998}. (d) Trade-offs in scenarios involving no ISA (scenario 1), duty-cycled communication (scenario 2), and compression with duty-cycled communication (scenario 3). [**Note to Annual Reviews: Sub-Figures adapted with permission from IEEE \cite{Chatterjee_DnT_2019, Cao_VLSI_2020}.  Chatterjee B and Sen S are authors of this article.**]}
\label{comm_en}
\end{figure}

\textbf{Figure \ref{comm_en}(b)} compares $\text{E}_\text{communication}$ with $\text{E}_\text{computation}$ for the same number of bits being computed, or communicated. State-of-the-art wireless transceiver implementations \cite{Mohseni_TBioCAS_2015} consume $\approx 10^4$ times more energy when compared to computational bit-switching in $45$ nm and $65$ nm technology nodes. This bottleneck-analysis directly signifies that intelligent computation within a sensor node (in-sensor analytics) can reduce the total energy consumption by enabling selective data transmission which decreases $\text{E}_\text{communication}$ at the cost of additional $\text{E}_\text{computation}$. \textbf{Figure \ref{comm_en}(c)} shows the trade-offs in implementing an analog computation unit vs digital computation unit in terms of the signal to noise ratio (SNR) requirement of the application \cite{Sarpeshkar_AD_1998}. For low SNR requirements, analog can perform much more power-efficient computations due to the use of inherent device properties for computation (for example, vector multiplication of transconductance with input voltage can generate a vector output current, and KCL can help adding those vector currents, without any bulky digital multiplier or adder). This also leads to lower number of devices in the implementation, resulting in lower power. However, analog suffers from accumulation of the effects of noise and mismatch over multiple stages, and hence it is beneficial only for applications that can tolerate low SNRs. Finally, \textbf{Figure \ref{comm_en}(d)} shows the trade-offs in 3 different scenarios. For scenario 1, no ISA is considered, and the sensed data is just transmitted out using continuous streaming, leading to high amount of power. In scenario 2, the communication is duty cycled, which results in lower data rates, lower power, but also lower information content. However, in scenario 3, by enabling compression and in-sensor storage, we can send more effective information per unit amount of power, at the cost of additional memory requirements.

\subsection{In-Sensor Analytics}
Based on the communication and computation energy trade-offs, the application at hand and the amount of resources available at resource-constrained bio-sensor node, partial or complete processing of the acquired data can take place in the sensor-node itself (for example, spike timing detection compression for an implanted neural node \cite{Sodagar_JSSC_2009, Sodagar_TBioCAS_2014, Mich_ISSCC_2020}). In this section, we discuss some of the common ISA techniques for bio-sensors, including anomaly/outlier detection and data compression, spike detection an spiking band power (SBP) calculation for neural sensors and machine learning based analytics.
\subsection{Examples of In-Sensor Analytics in Bio-Sensing}
\subsubsection{Anomaly Detection and Data Compression}
The anomaly detection methods can enable selective (and immediate) data transmission when an anomaly occurs in an otherwise normal sensor readout. As an example in healthcare, selective ECG data transmission with arrhythmia (anomaly) detection would ensure immediate notification with minimum communication cost. Data compression, on the other hand, would ensure that maximum amount of information between transmissions can be stored in a small amount of on-sensor memory. \cite{NV_MMADC_2015} showed a matrix-multiplying ADC (MM-ADC) in 130nm CMOS technology where the authors demonstrated ECG based cardiac arrhythmia detection, while \cite{AA_BioCAS_2017} showed arrhythmia detection with a time-based CS ADC.
\subsubsection{Spike detection for Neural Sensing}
In neural signal acquisition applications, a primary focus over the last few decades has been on increasing the number of recording channels. As the number of channels increase, the amount of data recorded keeps on increasing, and it becomes infeasible to communicate all the data from a neural implant to a nearby hub. However, most of the information in the neural signals reside in the spikes and hence detecting and communicating the occurrence and shape of the spikes is enough for most applications. Detection of spike occurrence is typically performed by traditional thresholding methods. Although the shape of the spike is not preserved, simple thresholding is still useful in the scenarios where only the occurrence of spikes is of interest \cite{Schwartz_ARNeuro_2004, Schwartz_TNSRE_2003}. Various implementations of thresholding for spike detection are shown in \cite{Wise_JSSC_2005} and \cite{Harrison_JSSC_2007}. Compression techniques are also demonstrated on neural spikes using optimized vector quantization methods \cite{SPIKE_OVQ_2011}, compressive sensing \cite{SPIKE_CS_2012, SPIKE_Allstot_CS_2012, Mohseni_CS_2018}, Wavelet Transform \cite{SPIKE_DWT_2007, SPIKE_DWT_2010}, Walsh-Hadamard Transform \cite{Sodagar_TBioCAS_2014} and Spiking Band Power (SBP) calculation \cite{Mich_ISSCC_2020, SBP_2016}. Wavelet Transform and Walsh-Hadamard Transform based techniques retain information on both the occurrence and shape of the spike.
\subsubsection{Learning-based Analytics} - For wearable scenarios where the bio-nodes have more resources than implantable scenarios, lightweight machine learning algorithms/tinyML can be implemented as a part of data analytics. \cite{Mohseni_Asilomar_2020} shows a transfer-learning based cuff-less blood pressure (BP) estimation technique using Photoplethysmography (PPG). The algorithm utilizes visibility graphs to create images from PPG signals with features related to the waveform morphology. Results demonstrate that the difference between the estimated and actual BP for systolic BP and diastolic BP are -0.080 $\pm$ 10.097 mmHg and 0.057 $\pm$ 4.814 mmHg, respectively. \cite{Mohseni_EMBC_2021} extends this work to a small-scale convolutional neural network (CNN), using a modified LeNet-5 architecture.

This section reinforces the need for incorporating certain processing capabilities in the bio-sensor nodes, based on their applications. There will be interesting trade-offs among sensing, processing and communication, which will be very different based on the application and the particular modalities chosen for these 3 functions. However, these trade-offs will be governed by the energy availability in the devices, which brings us to the next section, where we shall discuss the various techniques for powering these nodes.
\section{POWERING THE BIO-SENSORS: Energy Harvesting in a Wearable/ Implantable Device}
\label{powering}

\begin{figure}[htbp]
\includegraphics[width=0.95\columnwidth]{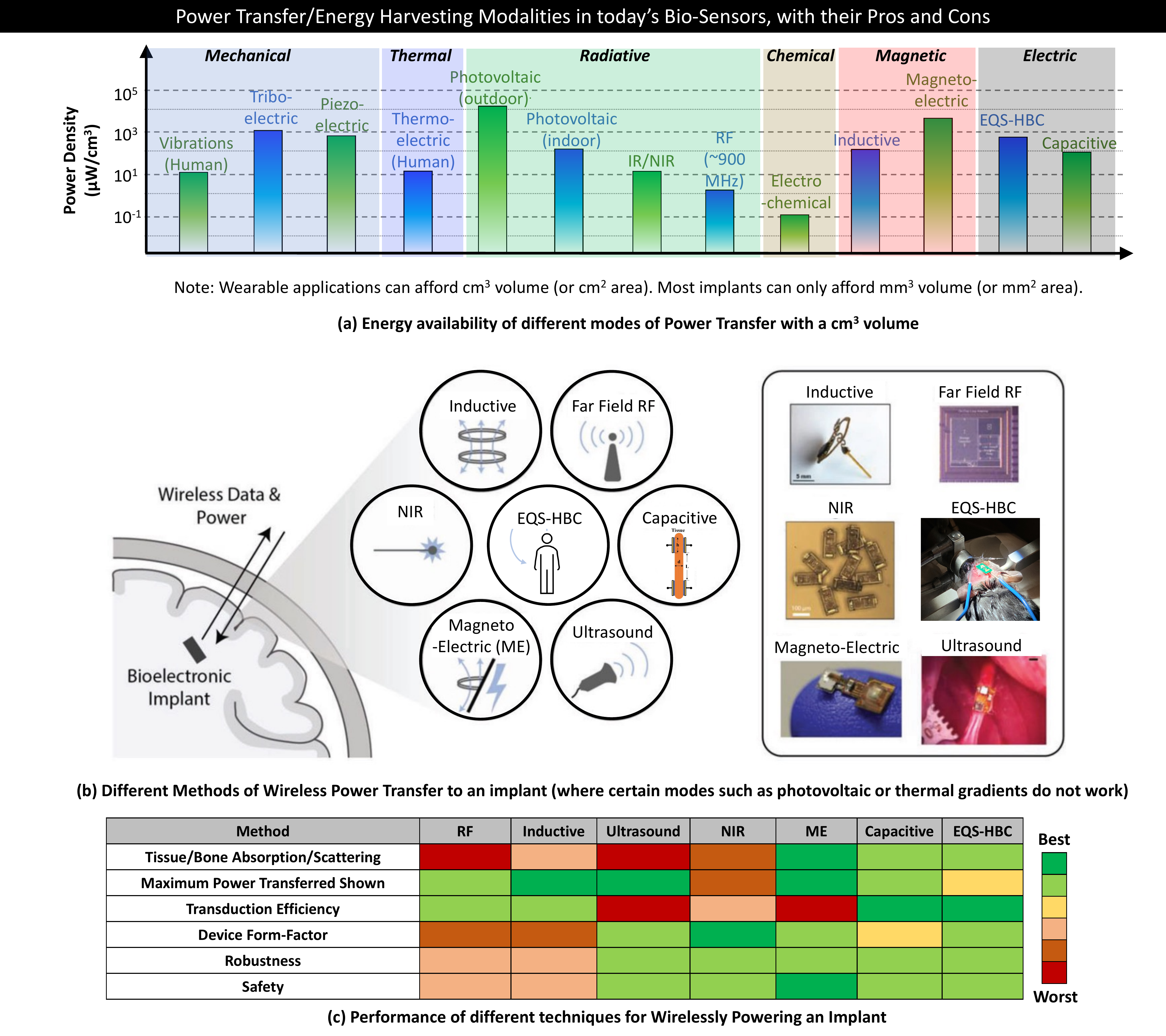}
\caption{(a) Available modalities of power transfer to a bio-sensor and their typical power densities; (b) As a subset, different modalities for powering an implant \cite{Singer_Health_2021} is shown; (c) Comparison of the power transfer modalities for an implant. [**Note to Annual Reviews: Part (b) is adapted with permission from IEEE \cite{Singer_Health_2021}. The publisher grants authors the right to reuse the figures without permission.**]}
\label{fig_Powering}
\end{figure}

In the last decade, various modalities of bio-node energy harvesting have been explored, including methods such as mechanical (vibration, tribo-electric and piezo-electric), thermal (thermo-electric), radiative (photovoltaic, IR, NIR, RF), chemical (electro-chemical), magnetic (inductive and magneto-electric) and electric (EQS-body-coupled and capacitive). The power densities for these techniques are shown in \textbf{Figure \ref{fig_Powering}(a)}.

As compared to wearable bio-sensors, the requirements of an implanted bio-sensor is much more stringent, due to the small form-factor, and lower amount of resources. For such scenarios, techniques including near-infrared (NIR) light \cite{Blaauw_ACS_2021, Blaauw_VLSI_2021}, ultrasound \cite{UCB_NeuroMethods_2015, UCB_Neuron_2016, UCB_ISSCC_2019, UCB_JSSC_2019, Verhelst_BioCAS_2019, Verhelst_TBioCAS_2019}, radio-frequency (RF)/Inductive \cite{Shepard_NatBio_2017, Mercier_NatE_2021, Mohseni_BioCAS_2018_1}, magnetoelectric \cite{Yang_ISSCC_2020, Robinson_NER_2021, Robinson_Neuron_2020}, capacitive \cite{Mohseni_ISSCC_2020, Mohseni_JSSC_2019, Mohseni_BioCAS_2018_2} and human body-coupled/electro-quasistatic \cite{JYoo_ISSCC_2020, JYoo_NatE_2021, Modak_CICC_2021, Chatterjee_VLSI_2021}  modes have been demonstrated in the recent past. The various modalities of powering an implant is shown in \textbf{Figure \ref{fig_Powering}(b)-(c)}. These modalities are also compared w.r.t. various parameters, such as (a) amount of tissue absorption/scattering, (b) maximum power transferred, (c) transduction efficiency, (d) form-factor of the device, (e) robustness and (f) safety. RF/Inductive methods suffers from tissue absorption, while ultrasound suffers from absorption in the bone (for example, in the skull for a brain implant). Hence, ultrasound methods require repeaters below the bone layer. In a similar manner, NIR also requires repeaters to improve both powering and communication. Magnetoelectric (ME) methods exhibit extremely low loss within human tissues, since the human body has a relative permeability of $\approx$ 1 at or below 10's of MHz frequencies. As a result, ME methods are also safer than EM or EQS mathods, and high magnetic fields can be applied across the human body so that the bio-nodes receive enough power. However, creation of this magnetic field requires significant energy, which is somewhat acceptable given that the magnetic field would be created from the body-worn hub which will have higher energy resources. Capacitive and EQS-HBC methods do not suffer from large tissue absorption or transduction losses. However, capacitive mode of powering requires parallel conductive plates to be placed on either side of the tissue (creating a a capacitor), and hence the form-factor of the device could increase. EQS-HBC, on the other hand, relies on creation of electric fields from the implant (bio-node) to the hub, and the signals traverses through the tissue following the equations for dipole coupling \cite{Chatterjee_VLSI_2021, Chatterjee_NatE_2022}, thereby reducing the received voltage as a function of the distance squared.

\subsection{Upcoming Methods of Powering a Wearable Device}
Power transfer using RF/Inductive methods has traditionally been applied for wearable nodes. Although the amount of power transferred using these methods can be in the range of 100's of $\mu$W or more for cm-scale devices, there are certain limitations related to device form-factors and the frequency to be used. Also, near-field power transfer is limited by distance, and far-field methods suffer from body-shadowing effects \cite{JYoo_NatE_2021}. Additionally, the orientation of the inductive coils at the TX and RX determine the amount of power transferred. Newer techniques for sending power through the human body is being explored recently \cite{JYoo_ISSCC_2020, JYoo_NatE_2021, Modak_CICC_2021} which can transfer power throughout the body (and any wearable placed anywhere on the body can pick it up) without any dependence on orientation.
\subsubsection{EQS Body-Coupled Powering}
\cite{JYoo_ISSCC_2020, JYoo_NatE_2021} showed the EQS powering techniques for the first time, and showed that body-coupled power transmission exhibits a path loss which is 30-70 dB lower than far-field RF in presence of body-shadowing. The wearable system works with an operating frequency of 40 MHz, and could recover 2 $\mu$W from a 1.2 mW source, placed 160 cm away on the body. However, the amount of power transferred reduces with the distance on the body.
\subsubsection{Resonant Whole-Body Powering}
\cite{Modak_CICC_2021, Modak_JSSC_2022} shows a resonant whole-body powering technique, by using frequencies around 1 MHz, which allows similar path loss throughout the body (a better EQS scenario). Measurements show $>$ 5 $\mu$W wearable-wearable power transfer from a $\approx$ 60 $\mu$W source ($\approx$ 8\% efficiency), independent of the device locations on the body. The system also shows a 2.19 $\mu$W communication system utilizing EQS-HBC, demonstrating the feasibility of communication with the harvested energy. \cite{Je_TBioCAS_2022} has recently developed an electrical model of such powering, and have shown that $>$1mW power can be transmitted for favorable scenarios including large devices and small distances. 

\subsection{Upcoming Methods of Powering an Implant}
The energy requirements of an implant is much more constrained than a wearable because of the small size, and inability to use and replace batteries. RF, inductive, ultrasound and optical methods of powering an implant suffers high amount of end-end-end loss (for example, RF gets absorbed in the tissue, while ultrasound is absorbed in the bones), and hence often requires repeaters for both powering and communication. Recently developed capacitive, bi-phasic EQS and magnetoelectric methods show promise in terms of the device form factors, low amount of tissue absorption and safety. 
\subsubsection{Capacitive Powering}
Capacitive powering techniques rely on power transfer using two differential plates - one set of plates are part of the implant, while the other set is wearable \cite{Mohseni_ISSCC_2020, Mohseni_JSSC_2019, Mohseni_BioCAS_2018_2}. With 20 mm $\times$ 20 mm capacitive patches with 3 mm thickness, and a 3 mm thick beef tissue sample in between, the maximum amout of power transfer was shown to be 12 mW, with 36 \% efficiency \cite{Mohseni_BioCAS_2018_2}. However, this method requires large capacitive plates and small distance of the implant from the surface of the body.  
\subsubsection{Bi-Phasic QBC}
\cite{Chatterjee_VLSI_2021} showed the bi-phasic mode of powering along with the communication. For neural nodes, a headphone-shaped body-worn hub can send power to the implant through differential excitation and pick-up. A maximum power of $>$ 1 $\mu$W is shown to be available at an implant 55 mm inside the brain with $>$ 35 V(rms) applied using the wearable hub. This method has low loss, and avoids transduction of signals to have high end-to-end efficiency. However, for powering, this method relies on creating large electric fields around the body, which could create safety concerns.
\subsubsection{Magnetoelectric (ME) Powering}
Utilizing the air-like permeability of the human body, magnetic fields can propagate through the tissue with almost no loss. Using this property, \cite{Yang_ISSCC_2020, Robinson_NER_2021, Robinson_Neuron_2020} has shown $\approx$ 400 $\mu$W of power transfer to an implant from an wearable that creates a 0.1 mT magnetic field at a distance of about 30 mm. For iso-energy density, this magnetic field is equivalent to a 300 kV/m electric field, which means that creation of such magnetic fields will consume significant energy. However, this could be justified since the magnetic field is created at a wearable hub, with access to better power sources.

From the discussion presented in the last 3 sections, we believe that EQS-HBC/capacitive modes for communication and ME methods for powering could become an optimum solution in near future for small implants such as wireless neural nodes. For wearable applications, on the other hand, body-coupled powering and HBC shows excellent promise toward developing extremely energy-efficient, perpetual bio-sensor nodes. 
\section{A NOTE ON SECURITY AND DATA PRIVACY}
\label{security}

Because of the inherent resource constraints in a small form-factor bio-node, advanced encryption techniques and high-overhead countermeasures for advanced attacks (such as side-channel attacks) often become infeasible to implement on the sensor node. However, standard encryption algorithms such as AES-256 has recently been shown to consume $<$ 200 nW power for $<$ 20 kbps data rates \cite{Maity_WRComm_2021} on a wearable sensor node.

\subsection{Threat Models}
\subsubsection{Side Channel Attacks (SCA):}
Even though the mathematical complexity of the key recovery algorithm for AES-256 is $2^{256}$ (which means a brute-force method would be able to break the encryption with a probability of 1/$2^{256}$), non-traditional techniques such as electromagnetic side channel attacks (EM-SCA) or power SCA can reduce the recovery complexity to only $2^{13}$, which can be broken within 50 seconds \cite{AES_2017}. Correlational Power Analysis (CPA) algorithms are used to break the AES key by looking at the fluctuations at the power lines during encryption, and hence poses lower risks for wearabale and implants, as the attacker needs physical access to the devices. On the other hand, Correlational EM Analysis (CEMA) can break the AES key by analyzing the EM radiation from the devices during encryption. CEMA-based SCA attacks pose a greater risk for bio-sensor security as the attack can be performed even from a distance by using a high-sensitivity receiver.
\subsubsection{Replay Attack/Mimicking a device:}
A malicious attacker can also impersonate a sensor-node and can replay/mimic the data transmitted by the original device, and hence create confusion at the receiver. If the malicious device transmits the data with enough power, it can also jam the intended transmissions, resulting in loss of important bio-physical information.

\subsection{Hardware Solutions}
Software-based masking techniques to prevent side-channel attacks require significant computational power. Hence, we shall focus more on low-overhead hardware solutions for both SCA and replay attacks.
\subsubsection{SCA Countermeasures:}
\cite{Das_DnT_2021} shows a white-box modeling and signature attenuation countermeasure for EM/Power SCA, based on techniques such as current-domain signature attenuation - CDSA \cite{Das_TCAS_2018, Das_JSSC_2021, Das_ISSCC_2020} that helps in reducing the signatures utilized in Power SCA, along with local EM signature suppression through low-level metal routing \cite{Das_JSSC_2021, Das_Stellar_2019}.
 
 Current-domain signature attenuation (CDSA) uses a constant current source to supply the AES current. This reduces the correlated fluctuations on the power lines during encryption and improves the Figure of merit (FoM) for CPA countermeasure. On the other hand, local EM signature suppression is implemented by routing the encryption hardware (the crypto engine) using lower-level metals in a complementary metal-oxide-semiconductor (CMOS) process. This ensures that the higher level thicker metals in the CMOS process do not carry correlated currents during encryption, which would have otherwise leaked the critical information in the form of EM signals. Using these techniques, \cite{Das_JSSC_2021} achieves $>$ 33 $\times$ SCA security improvement w.r.t. the state-of-the-art. \cite{Ghosh_ISSCC_2021} demonstrates a digital-friendly implementation of CDSA with low-level metal routing for fast-time-to-market applications. 
 
 \subsubsection{Countermeasures against data mimicking:} Physically unclonable function (PUF) based techniques are used to generate a device-specific signature, utilizing the inherent manufacturing process variations. However, this usually is done in the digital domain, which requires additional hardware and processing steps. \cite{Chatterjee_JIoT_2019} proposed a new type of PUF for resource-constrained IoT and bio-sensor nodes, that utilizes the intrinsic analog and RF properties of any transmitter node. Every transmitter inherently possesses analog/RF non-idealities on top of an intended digital signal (or, the mimicked digital data in case of a malicious device) due to certain variations in the manufacturing process. By utilizing the property that these non-idealities are different for every transmitter, an in-situ machine learning hardware at the resource-rich receiver can detect a particular transmitter through analysis of these non-idealities. A malicious transmitter that mimics (or alters) someone else's data can also be identified separately from the intended transmitters. This system was motivated by the unique voice signatures in human-to-human communication, which our brain uses to map the identity of a person with their voice.
 
 There is undoubtedly a strong demand of incorporating security features during design-time for the resource-constrained, small bio-sensor nodes to protect personal and health-related information, emphasizing the need for further exploration of low-cost, lightweight hardware security primitives. As a result, this is fast becoming one of the major areas of current research in bio-sensors.

\section{CONCLUSION: The Vision for the Future}
\label{Conclude}
Bio-sensors are different from traditional wireless sensor networks, considering the specific challenges in wearable/implantable applications, including resource (energy/computation/memory) availability , security, and powering. In this review, we identify the primary design objective of such a connected bio-sensor node, which is to minimize the energy consumption per unit amount of information. We have focused on a broad analysis of the constituent units of such a sensor, in terms of Sensing, Processing, Communication, and Powering, all of which needs to be designed optimally for holistic, system-level resource optimization for constrained bio-sensors. The numerous challenges in the form of system-level adaptive control, reliability, security, latency  and powering limitations are also discussed, indicating future research directions toward smart, secure and connected Bioelectronic Medicine, as well as IoB in general, that has the promise to enrich human lives by leveraging the progress in the various technologies discussed throughout the paper.

\section*{DISCLOSURE STATEMENT}
The authors are not aware of any affiliations, memberships, funding, or financial holdings that might be perceived as affecting the objectivity of this review. 

\section*{ACKNOWLEDGMENTS}
This work is supported in part by the National Science Foundation (NSF) Career Award, grant \#1944602.



\begin{thebibliography}{00}


\bibitem{Das_SciRep_2019}
Das D, Maity S, Chatterjee B, Sen S. 2019. Enabling Covert Body Area Network using Electro-Quasistatic Human Body Communication. {\it Scientific Reports} 9:4160

\bibitem{Chatterjee_DnT_2019}
Chatterjee B, Cao N, Raychowdhury A, Sen S. 2019. Context-Aware Intelligence in Resource-Constrained IoT Nodes: Opportunities and Challenges. {\it IEEE Design \& Test} 36(2):7--40

\bibitem{Cao_VLSI_2020}
Cao N, Chatterjee B, Gong M, Chang M, Sen S, Raychowdhury A. 2020. A 65nm Image Processing SoC Supporting Multiple DNN Models and Real-Time Computation-Communication Trade-Off Via Actor-Critical Neuro-Controller. {\it 2020 IEEE Symposium on VLSI Circuits - (VLSI)}:1--2.

\bibitem{Chatterjee_JIoT_2021}
Chatterjee B, Seo DH, Chakraborty S, Avlani S, Jiang X, Zhang H, Abdallah M, Raghunathan N, Mousoulis C, Shakouri S, Bagchi S, Peroulis D, Sen S. 2021. Context-Aware Collaborative Intelligence With Spatio-Temporal In-Sensor-Analytics for Efficient Communication in a Large-Area IoT Testbed. {\it IEEE Internet of Things Journal} 8(8):6800--14

\bibitem{DHS_FDA_2019}
U.S. Cybersecurity and Infrastructure Security Agency. 2019. ICS Medical Advisory (ICSMA-19-080-01). https://us-cert.cisa.gov/ics/advisories/ICSMA-19-080-01

\bibitem{Spectrum_2019}
Waltz E, IEEE Spectrum. 2019. Can ``Internet-of-Body" Thwart Cyber Attacks on Implanted Medical Devices? https://spectrum.ieee.org/thwart-cyber-attacks-on-implanted-medical-devices

\bibitem{Sen_Spectrum_2020}
Sen S, Maity S, Das D. 2020. The Body is the Network: To Safeguard Sensitive Data, Turn Flesh and Tissue into a Secure Wireless Channel. {\it IEEE Spectrum}, 57(12):44--49


\bibitem{Maity_WRComm_2021}
Maity S, Modak N, Yang D, Nath M, Avlani S, Das D, Danial J, Mehrotra P, Sen S. 2021. Sub-$\mu$WRComm: 415-nW 1–10-kb/s Physically and Mathematically Secure Electro-Quasi-Static HBC Node for Authentication and Medical Applications. {\it IEEE Journal of Solid-State Circuits} 56(3):788--802


\bibitem{Chatterjee_VLSI_2021}
Chatterjee B, Gaurav K, Nath M, Xiao S, Modak N, Das D, Krishna J, Sen S. 2021. A 1.15$\mu$W 5.54mm$^3$ Implant with a Bidirectional Neural Sensor and Stimulator SoC utilizing Bi-Phasic Quasi-static Brain Communication achieving 6kbps-10Mbps Uplink with Compressive Sensing and RO-PUF based Collision Avoidance. {\it 2021 Symposium on VLSI Circuits}: 1-2

\bibitem{Chatterjee_NatE_2022} 
Chatterjee B, Nath M, Gaurav K, Xiao S, Krishna J, Sen S. 2022. Bi-Phasic Quasistatic Brain Communication for Fully Untethered Connected Brain Implants. {\it arXiv}:2205.08540


\bibitem{IoB_2020}
Rand Corporation. 2020. What Is the Internet of Bodies? https://www.rand.org/multimedia/ video/2020/10/29/what-is-the-internet-of-bodies.html

\bibitem{CIoB_2021}
Purdue University. 2021. Center for Internet of Bodies. https://engineering.purdue.edu/C-IoB


\bibitem{Wiki_Li_2021}
Wikipedia. Lithium-ion battery. 2021. https://en.wikipedia.org/wiki/Lithium-ion\_battery\

\bibitem{Sen_DAC_2016}
Sen S. 2016. Invited: Context-aware Energy-efficient Communication for IoT Sensor Nodes. {\it 2016 53nd ACM/EDAC/IEEE Design Automation Conference (DAC)}: 1--6

\bibitem{Candes_CS_2005} 
Candes EJ, Tao T. 2005. Decoding by Linear Programming. {\it IEEE Transactions on Information Theory} 51(12):4203--15

\bibitem{Donoho_CS_2006}
Donoho DL. 2006. Compressed Sensing. {\it IEEE Transactions on Information Theory} 52(4):1289-1306

\bibitem{Jha_IoT_2017}
Mosenia A, Jha NK. 2017. A Comprehensive Study of Security of Internet-of-Things. {\it IEEE Transactions on Emerging Topics in Computing} 5(4):586-602

\bibitem{Das_DnT_2021}
Das D, Ghosh S, Raychowdhury A, Sen S. 2021. EM/Power Side-Channel Attack: White-Box Modeling and Signature Attenuation Countermeasures, {\it IEEE Design \& Test} 38(3): 67--75

\bibitem{Das_TCAS_2018}
Das D, Maity S, Nasir SB, Ghosh S, Raychowdhury A, Sen S. 2018. ASNI: Attenuated Signature Noise Injection for Low-Overhead Power Side-Channel Attack Immunity. {\it IEEE Transactions on Circuits and Systems I: Regular Papers} 65(10):3300--11

\bibitem{Das_JSSC_2021} 
Das D, Danial J, Golder A, Modak N, Maity S, Chatterjee B, Seo DH, Chang M, Varna A, Krishnamurthy H, Matthew S, Ghosh S, Raychowdhury A, Sen S. 2021. EM and Power SCA-Resilient AES-256 Through $>$350× Current-Domain Signature Attenuation and Local Lower Metal Routing. {\it IEEE Journal of Solid-State Circuits} 56(1):136--50

\bibitem{Das_Stellar_2019}
Das D, Nath M, Chatterjee B, Ghosh S, Sen S. 2019. STELLAR: A Generic EM Side-Channel Attack Protection through Ground-Up Root-cause Analysis. {\it 2019 IEEE International Symposium on Hardware Oriented Security and Trust (HOST)}: 11--20

\bibitem{Chatterjee_JIoT_2019}
Chatterjee B, Das D, Maity S, Sen S. 2019. RF-PUF: Enhancing IoT Security Through Authentication of Wireless Nodes Using In-Situ Machine Learning. {\it IEEE Internet of Things Journal} 6(1):388--98

\bibitem{Bari_IMS_2021}
Bari MF, Chatterjee B, Sivanesan K, Yang LL, Sen S. 2021. High Accuracy RF-PUF for EM Security through Physical Feature Assistance using Public Wi-Fi Dataset. {IEEE MTT-S International Microwave Symposium (IMS)}:108--11

\bibitem{Blaauw_ACS_2021}
Moon E, Barrow M, Lim J, Lee J, Nason SR, Costello J, Kim HS, Chestek CA, Jang T, Blaauw D, Phillips JD. 2021. Bridging the ``Last Millimeter" Gap of Brain-Machine Interfaces via Near-Infrared Wireless Power Transfer and Data Communications. {\it ACS Photonics} 8(5):1430--38

\bibitem{Blaauw_VLSI_2021} 
Lim J, Lee J, Moon E, Barrow M, Atzeni G, Letner J, Costello J, Nason SR, Patel PR, Patil PG, Kim HS, Chestek CA, Phillips JD, Blaauw D, Sylvester D, Jang T. 2021. A Light Tolerant Neural Recording IC for Near-Infrared-Powered Free Floating Motes. {\it 2021 Symposium on VLSI Circuits}: 1--2

\bibitem{UCB_NeuroMethods_2015}
Seo D, Carmena JM, Rabaey JM, Maharbiz MM, Alon E. 2015. Model Validation of Untethered, Ultrasonic Neural Dust Motes for Cortical Recording. {\it Journal of Neuroscience Methods} 244:114--22

\bibitem{UCB_Neuron_2016}
Seo D, Neely RM, Shen K, Singhal U, Alon E, Rabaey JM, Carmena JM, Maharbiz MM. 2016. Wireless Recording in the Peripheral Nervous System with Ultrasonic Neural Dust. {\it Neuron} 91(3):529--39

\bibitem{UCB_ISSCC_2019}
Ghanbari MM, Piech DK, Shen K, Alamouti SF, Yalcin C, Johnson BC, Carmena JM, Maharbiz MM, Muller R. 2019. A 0.8mm$^3$ Ultrasonic Implantable Wireless Neural Recording System With Linear AM Backscattering. {\it 2019 IEEE International Solid- State Circuits Conference - (ISSCC)}:284--86

\bibitem{UCB_JSSC_2019}
Ghanbari MM, Piech DK, Shen K, Alamouti SF, Yalcin C, Johnson BC, Carmena JM, Maharbiz MM, Muller R. 2019. A Sub-mm$^3$ Ultrasonic Free-Floating Implant for Multi-Mote Neural Recording. {\it IEEE Journal of Solid-State Circuits} 54(11):3017--30

\bibitem{Verhelst_BioCAS_2019} 
Bos T, Dehaene W, Verhelst M. 2019. Ultrasound In-Body Communication with OFDM through Multipath Realistic Channels. {\it 2019 IEEE Biomedical Circuits and Systems Conference (BioCAS)}:1-4

\bibitem{Verhelst_TBioCAS_2019}
Bos T, Jiang W, D’hooge J, Verhelst M, Dehaene W. 2019. Enabling Ultrasound In-Body Communication: FIR Channel Models and QAM Experiments, {\it IEEE Transactions on Biomedical Circuits and Systems} 13(1):135--144.

\bibitem{Ren_SciAdv_2021}
Ren W, Sun Y, Zhao D, Aili A, Zhang S, Shi C, Zhang J, Geng H, Zhang J, Zhang L, Xiao J, Yang R. 2021. High-Performance Wearable Thermoelectric Generator with Self-Healing, Recycling, and Lego-like Reconfiguring Capabilities. {\it Science Advances} 7(7):eabe0586

\bibitem{Shepard_NatBio_2017}
Thimot J, Shepard KL. 2017. Bioelectronic Devices: Wirelessly powered implants. {\it Nature Biomedical Engineering} 1:0051

\bibitem{Mercier_NatE_2021}
Lee J, Leung V, Lee AH, Huang J, Asbeck P, Mercier PP, Shellhammer S, Larson L, Laiwalla F, Nurmikko, A. 2021. Neural Recording and Stimulation using Wireless Networks of Microimplants. {\it Nature Electronics} 4:604--14

\bibitem{Mohseni_BioCAS_2018_1} 
Vitale NR, Azin M, Mohseni P. 2018. A Bluetooth Low Energy (BLE)-enabled Wireless Link for Bidirectional Communications with a Neural Microsystem. {\it 2018 IEEE Biomedical Circuits and Systems Conference (BioCAS)}:1-4

\bibitem{Yang_ISSCC_2020}
Yu Z, Chen JC, Avants BW, He Y, Singer A, Robinson JT, Yang K. 2020.  An 8.2 mm$^3$ Implantable Neurostimulator with Magnetoelectric Power and Data Transfer. {\it 2020 IEEE International Solid-State Circuits Conference-(ISSCC)}:510--12

\bibitem{Robinson_NER_2021}
Alrashdan FT, Chen JC, Singer A, Avants BW, Yang K, Robinson JT. 2021. Wearable Wireless Power Systems for ‘ME-BIT’ Magnetoelectric-Powered Bio-Nodes. {\it Journal of Neural Engineering} 18(4):045011

\bibitem{Robinson_Neuron_2020}
Singer A, Dutta S, Lewis E, Chen Z, Chen JC, Verma N, Avants BW, Feldman AK, O’Malley J, Beierlein M, Kemere C, Robinson JT. 2020. Magnetoelectric Materials for Miniature, Wireless Neural Stimulation at Therapeutic Frequencies. {\it Neuron} 107(4):631--43

\bibitem{Mohseni_ISSCC_2020}
Marefat F, Erfani R, Kilgore KL, Mohseni P. 2020.  A 280$\mu$W 108dB DR Readout IC with Wireless Capacitive Powering Using a Dual-Output Regulating Rectifier for Implantable PPG Recording. {\it 2020 IEEE International Solid-State Circuits Conference-(ISSCC)}:412--14

\bibitem{Mohseni_JSSC_2019} 
Erfani R, Marefat F, Nag S, Mohseni P. 2019.  A 1–10-MHz Frequency-Aware CMOS Active Rectifier with Dual-loop Adaptive Delay Compensation and $>$ 230-mW Output Power for Capacitively Powered Biomedical Implants. {\it IEEE Journal of Solid-State Circuits} 55(3):756--66

\bibitem{Mohseni_BioCAS_2018_2}
Koruprolu A, Nag S, Erfani R, Mohseni P. 2018.  Capacitive Wireless Power and Data Transfer for Implantable Medical Devices. {\it 2018 IEEE Biomedical Circuits and Systems Conference (BioCAS)}:1--4

\bibitem{JYoo_ISSCC_2020}
Li J, Dong Y, Park JH, Lin L, Tang T, Zhang M, Wu H, Zhang L, Tan JSY, Yoo J. 2020. Human-Body-Coupled Power-Delivery and Ambient-Energy-Harvesting ICs for a Full-Body-Area Power Sustainability. {\it 2020 IEEE International Solid- State Circuits Conference - (ISSCC)}:514--516

\bibitem{JYoo_NatE_2021}
Li J, Dong Y, Park JH, Yoo J. 2021. Body-coupled Power Transmission and Energy Harvesting. {\it Nature Electronics} 5:530--38

\bibitem{Modak_CICC_2021}
Modak N, Das D, Nath M, Chatterjee B, Gaurav K, Maity S, Sen S. 2021. A 65nm Resonant Electro-Quasistatic 5-240uW Human Whole-Body Powering and 2.19$\mu$W Communication SoC with Automatic Maximum Resonant Power Tracking. {\it 2021 IEEE Custom Integrated Circuits Conference (CICC)}: 1--2

\bibitem{Modak_JSSC_2022} 
Modak N, Das D, Nath M, Chatterjee B, Gaurav K, Maity S, Sen S. 2022. EQS Res-HBC: A 65-nm Electro-Quasistatic Resonant 5–240 $\mu$W Human Whole-Body Powering and 2.19 $\mu$W Communication SoC With Automatic Maximum Resonant Power Tracking. {\it IEEE Journal of Solid-State Circuits} 57(3): 831--44

\bibitem{Je_TBioCAS_2022}
Cho H, Suh J-H, Kim C, Ha S, Je M. 2022. An Intra-Body Power Transfer System With  $>$ 1-mW Power Delivered to the Load and 3.3-V DC Output At 160-Cm of On-Body Distance. {\it IEEE Transactions on Biomedical Circuits and Systems}

\bibitem{Kilgore_TBioCAS_2021}
Makowski NS, Campean A, Lambrecht JM, Buckett JR, Coburn JD, Hart RL, Miller ME, Montague FW, Crish T, Fu MJ, Kilgore KL, Peckham PH, Smith B. 2021. Design and Testing of Stimulation and Myoelectric Recording Modules in an Implanted Distributed Neuroprosthetic System.  {\it IEEE Transactions on Biomedical Circuits and Systems} 15(2):281--93

\bibitem{CMS_NHED_2019}
U.S. Centers for Medicare and Medicaid Services. 2019. National Health Expenditure Data. https://www.cms.gov/Research-Statistics-Data-and-Systems/Statistics-Trends-and-Reports/NationalHealthExpendData/NationalHealthAccountsHistorical


\bibitem{SAR_ADC_2019}
Zhang M, Chan C, Zhu Y, Martins RP. 2019. A 0.6-V 13-bit 20-MS/s Two-Step TDC-Assisted SAR ADC With PVT Tracking and Speed-Enhanced Techniques. {\it IEEE Journal of Solid-State Circuits} 54(12):3396--3409

\bibitem{Chatterjee_RFIC_2020} 
Chatterjee B, Srivastava A, Seo DH, Yang D, Sen S. 2020. A Context-aware Reconfigurable Transmitter with 2.24 pJ/bit, 802.15.6 NB-HBC and 4.93 pJ/bit, 400.9 MHz MedRadio Modes with 33.6\% Transmit Efficiency. {\it 2020 IEEE Radio Frequency Integrated Circuits Symposium (RFIC)}: 75--78.


\bibitem{Singer_Health_2021}
Singer A, Robinson JT. 2021. Wireless Power Delivery Techniques for Miniature Implantable Bioelectronics. {\it Advanced Healthcare Materials} 10(17):2100664





\bibitem{Yazicioglu_2007}
Yazicioglu R F, Merken P, Puers R, Van Hoof C. 2007. A 60 $\mu$W 60 nV/$\sqrt{\text{Hz}}$ Readout Front-End for Portable Biopotential Acquisition Systems. {\it IEEE Journal of Solid-State Circuits} 42(5):1100--10

\bibitem{Enz_1996}
Enz C C, Temes G C. 1996. Circuit Techniques for reducing the effects of Op-Amp Imperfections: Autozeroing, Correlated Double Sampling, and Chopper Stabilization. {\it Proceedings of the IEEE} 84(11): 1584--14

\bibitem{Denison_2007}
Denison T, Consoer K, Santa W, Avestruz A T, Cooley J, Kelly A. 2007. A 2 $\mu$W  100 nV/$\sqrt{\text{Hz}}$ Chopper-Stabilized Instrumentation Amplifier for Chronic Measurement of Neural Field Potentials. {\it IEEE Journal of Solid-State Circuits} 42(12):2934--45

\bibitem{Markovic_2017} 
Chandrakumar H, Markovi\'{c} D. 2017. A High Dynamic-Range Neural Recording Chopper Amplifier for Simultaneous Neural Recording and Stimulation. {\it IEEE Journal of Solid-State Circuits} 52(3):645--56

\bibitem{Mehrotra_Sensors_2019}
Mehrotra P, Chatterjee B, Sen S. 2019. EM-Wave Biosensors: A Review of RF, Microwave, mm-Wave and Optical Sensing. {\it Sensors} 19(5):1013

\bibitem{Murmann_2022}
Boris Murmann. 2022. ADC Performance Survey 1997-2022. https://web.stanford.edu/~murmann/adcsurvey.html

\bibitem{Chatterjee_CICC_2019}
Chatterjee B, Mousoulis C, Maity S, Kumar A, Scott S, Valentino D, Peroulis D, Sen S. 2019. A Wearable Real-Time CMOS Dosimeter With Integrated Zero-Bias Floating Gate Sensor and an 861-nW 18-Bit Energy-Resolution Scalable Time-Based Radiation to Digital Converter. {\it 2019 IEEE Custom Integrated Circuits Conference (CICC)}:1--4

\bibitem{Chatterjee_JSSC_2020}
Chatterjee B, Mousoulis C, Seo DH, Kumar A, Maity S, Scott S, Valentino D, Morisette D, Peroulis D, Sen S. 2020. A Wearable Real-Time CMOS Dosimeter With Integrated Zero-Bias Floating Gate Sensor and an 861-nW 18-Bit Energy-Resolution Scalable Time-Based Radiation to Digital Converter. {\it IEEE Journal of Solid-State Circuits} 55(3):650--65


\bibitem{CS_MRI_2007} 
Lustig M, Donoho D, Pauly JM. 2007. Sparse MRI: The Application of Compressed Sensing for Rapid MR Imaging. {\it Magnetic Resonance in Medicine} 58(6):1182--95

\bibitem{AA_TCASII_2017}
Anvesha A, Xu S, Romberg J, Raychowdhury A. 2017. A 130nm 165nJ/frame Compressed-Domain Smashed-Filter Based Mixed-Signal Classifier for ``In-sensor" Analytics in Smart Cameras. {\it Transactions on Circuits and Systems II: Express Briefs} 65(3):296--300

\bibitem{NV_MMADC_2015}
Zhang J, Wang Z, Verma N. 2015. A matrix-multiplying ADC implementing a machine-learning classifier directly with data conversion. {\it 2015 IEEE International Solid-State Circuits Conference-(ISSCC)} 7(4):1--3

\bibitem{AA_BioCAS_2017}
Anvesha A, Xu S, Romberg J, Raychowdhury A. 2017. A 65nm Compressive-sensing Time-based ADC with Embedded Classification and INL-aware Training for Arrhythmia Detection. {\it 2017 IEEE Biomedical Circuits and Systems Conference (BioCAS)}: 1--4

\bibitem{Gaurav_CICC_2021}
Gaurav K, Chatterjee B, Sen S. 2021. A 16 pJ/bit 0.1-15Mbps Compressive Sensing IC with on-chip DWT Sparsifier for Audio Signals. {\it 2021 IEEE Custom Integrated Circuits Conference (CICC)}: 1--2

\bibitem{Gaurav_JSSC_2022} 
Gaurav K, Chatterjee B, Sen S. 2021. CS-Audio: A 16 pJ/b 0.1–15 Mbps Compressive Sensing IC With DWT Sparsifier for Audio-AR. {\it IEEE Journal of Solid-State Circuits} 57(7): 2220--35

\bibitem{CS_Survey_2013}
Qaisar S, Bilal RM, Iqbal W, Naureen M, Lee S. 2013. Compressive Sensing: From Theory to Applications, A Survey. {\it Journal of Communications and Networks} 15(5):443--56

\bibitem{CS_Survey_2018}
Hamza D, Abbes A, Faycal B. 2018. Compressive Sensing-Based IoT Applications: A Review. {\it Journal of Sensor and Actuator Networks} 7(4):45


\bibitem{Verhelst_ESSIRC_2018}
Roose JD, Xin H, Andraud M, Harpe PJA. Verhelst M. 2018. Flexible and Self-Adaptive Sense-and-Compress for Sub-MicroWatt Always-on Sensory Recording. {\it 2018 IEEE 44th European Solid State Circuits Conference (ESSCIRC)}: 282--85

\bibitem{Trakimas_TCAS_2011}
Trakimas M, Sonkusale SR. 2011. An Adaptive Resolution Asynchronous ADC Architecture for Data Compression in Energy Constrained Sensing Applications. {\it IEEE Transactions on Circuits and Systems I: Regular Papers} 58(5):921--34

\bibitem{Martins_TVLSI_2017} 
Ieong C, Li M, Law M, Mak P, Vai MI, Martins RP. 2017. A 0.45 V 147–375 nW ECG Compression Processor With Wavelet Shrinkage and Adaptive Temporal Decimation Architectures. {\it IEEE Transactions on Very Large Scale Integration (VLSI) Systems} 25(4):1307--19


\bibitem{JoshuaSmith_USENIX_2018}
Naderiparizi S, Hessar M, Talla V, Gollakota S, Smith JR. 2018. Towards Battery-Free HD Video Streaming. {\it 2018 15th USENIX Symposium on Networked Systems Design and Implementation (NSDI)}: 233--47

\bibitem{Chatterjee_ESSCIRC_2022}
Chatterjee B, Gaurav K, Xiao S, Barik G, Krishna J, Sen S. 2022. A 1.8$\mu$W 5.5mm$^3$ ADC-less Neural Implant SoC utilizing 13.2pJ/Sample Time-domain Bi-phasic Quasi-static Brain Communication with Direct Analog to Time Conversion. {\it 2022 European Conference on Solid-State Circuits (ESSCIRC)}

\bibitem{CollabSense_2011}
Li W, Bao J, Shen W. 2011. Collaborative Wireless Sensor Networks: A Survey. {\it 2011 IEEE International Conference on Systems, Man, and Cybernetics)}: 2614--19

\bibitem{CollabSense_2018}
Buehrer RM, Wymeersch H, Vaghefi RM. 2018. Collaborative Sensor Network Localization: Algorithms and Practical Issues. {\it Proceedings of the IEEE} 106(6):1089-1114



\bibitem{Chang_JETCAS_2018} 
Chang G, Maity S, Chatterjee B, Sen S. 2018. A MedRadio Receiver Front-End With Wide Energy-Quality Scalability Through Circuit and Architecture-Level Reconfigurations. {\it IEEE Journal on Emerging and Selected Topics in Circuits and Systems} 8(3):369--378

\bibitem{Cao_JSSC_2022}
Cao N,  Chatterjee B, Liu, Cheng B, Gong M, Liu J, Chang M, Sen S, Raychowdhury A. 2022. A 65 nm Wireless Image SoC Supporting On-Chip DNN Optimization and Real-Time Computation-Communication Trade-Off via Actor-Critical Neuro-Controller. {\it IEEE Journal of Solid-State Circuits} 57(8):2545--59

\bibitem{Maity_BodyWire_2019}
Maity S, Chatterjee B, Chang G, Sen S. 2019. BodyWire: A 6.3-pJ/b 30-Mb/s -30-dB SIR-Tolerant Broadband Interference-Robust Human Body Communication Transceiver Using Time Domain Interference Rejection. {\it IEEE Journal of Solid-State Circuits} 54(10):2892--2906

\bibitem{Mondal_RFIC_2020}
Mondal S, Hall D A. 2020. A 67-$\mu$W Ultra-Low Power PVT-Robust MedRadio Transmitter. {\it 2020 IEEE Radio Frequency Integrated Circuits Symposium (RFIC)}: 327-330

\bibitem{Lucev_HBC_2012}
Lucev \v{Z}, Krois I, Cifrek M. 2012. A Capacitive Intrabody Communication Channel from 100 kHz to 100 MHz.  {\it IEEE Transactions on Instrumentation and Measurement} 61(12):3280--89

\bibitem{Bae_HBC_2012} 
Bae J, Cho H, Song K, Lee H, Yoo HJ 2012. The Signal Transmission Mechanism on the Surface of Human Body for Body Channel Communication. {\it IEEE Transactions on Microwave Theory and Techniques} 60(3):582-93

\bibitem{Park_TBME_2017}
Park J, Garudadri H, Mercier PP. 2017. Channel Modeling of Miniaturized Battery-Powered Capacitive Human Body Communication Systems. {\it IEEE Transactions on Biomedical Engineering} 64(2):452--62

\bibitem{Maity_TBME_2019}
Maity S, He M, Nath M, Das D, Chatterjee B, Sen S. 2019. Bio-Physical Modeling, Characterization, and Optimization of Electro-Quasistatic Human Body Communication. {\it IEEE Transactions on Biomedical Engineering} 66(6):1791--1802

\bibitem{Nath_TCAS2_2020}
Nath M, Maity S, Sen S. 2020. Toward Understanding the Return Path Capacitance in Capacitive Human Body Communication. {\it IEEE Transactions on Circuits and Systems II: Express Briefs} 67(10):1879--83

\bibitem{Maity_EMBC_2018}
Maity S, Das D, Chatterjee B, Sen S. 2018. Characterization and Classification of Human Body Channel as a function of Excitation and Termination Modalities. {\it 2018 Annual International Conference of the IEEE Engineering in Medicine and Biology Society (EMBC)}: 3754--57

\bibitem{Chatterjee_ISSCC_2022} 
Chatterjee B, Datta A, Nath M, Gaurav K, Modak N, Sen S. 2022. A 65nm 63.3$\mu$W 15Mbps Transceiver with Switched-Capacitor Adiabatic Signaling and Combinatorial-Pulse-Position Modulation for Body-Worn Video-Sensing AR Nodes. {\it 2022 IEEE International Solid- State Circuits Conference (ISSCC)}: 276--78

\bibitem{Modak_TBME_2021}
Modak N, Nath M, Chatterjee B, Maity S, Sen S. 2021. Bio-Physical Modeling of Galvanic Human Body Communication in Electro-Quasistatic Regime. {\it bioRxiv} 2020.11.23.394395

\bibitem{Mag_EMBC_2015}
Park J, Mercier PP. 2015. Magnetic Human Body Communication. {\it 2015 Annual International Conference of the IEEE Engineering in Medicine and Biology Society (EMBC)}: 1841--44

\bibitem{Mag_ISSCC_2019}
Park J, Mercier PP. 2019. A Sub-40$\mu$W 5 Mb/s Magnetic Human Body Communication Transceiver Demonstrating Trans-Body Delivery of High-Fidelity Audio to a Wearable In-Ear Headphone. {\it 2019 IEEE International Solid-State Circuits Conference-(ISSCC)}: 1841--44

\bibitem{Mag_TBME_2021}
Nath M, Ulvog AK, Weigand S, Sen S. 2021. Understanding The Role of Magnetic and
Magneto-Quasistatic Fields in Human Body Communication. {\it arXiv}: 2011.00125

\bibitem{Tochou_JSSC_2022} 
Tochou G, Benarrouch R, Gaidioz D, Cathelin A, Frapp\'{e} A, Kaiser A, Rabaey J. 2022. A Sub-100-$\mu$W 0.1-to-27-Mb/s Pulse-Based Digital Transmitter for the Human Intranet in 28-nm FD-SOI CMOS. {\it IEEE Journal of Solid-State Circuits} 57(5): 1409-20

\bibitem{Lee_JSSC_2019}
Lee M-C, Karimi-Bidhendi A, Malekzadeh-Arasteh O, Wang PT, Do AH, Nenadic Z, Heydari P. 2019. A CMOS MedRadio Transceiver With Supply-Modulated Power Saving Technique for an Implantable Brain–Machine Interface System. {\it IEEE Journal of Solid-State Circuits} 54(6): 1541--52.

\bibitem{IO_ISSCC_2014}
Jaussi JE, Balamurugan G, Hyvonen S, Hsueh TC, Musah T, Keskin G, Shekhar S, Kennedy J, Sen S, Inti R, Mansuri M, Leddige M, Horine B, Roberts C, Mooney R, Casper B. 2014. A 205mW 32Gb/s 3-Tap FFE/6-tap DFE bidirectional serial link in 22nm CMOS. {\it 2014 IEEE International Solid-State Circuits Conference-(ISSCC)}:440--41


\bibitem{Hsueh_ISSCC_2014}
Hsueh TC, Balamurugan G, Jaussi JE, Hyvonen S, Kennedy J, Keskin G, Musah T, Shekhar S, Inti R, Sen S, Mansuri M, Roberts C, Casper B. 2014. A 25.6Gb/s Differential and DDR4/GDDR5 Dual-mode Transmitter with Digital Clock Calibration in 22nm CMOS. {\it 2014 IEEE International Solid-State Circuits Conference-(ISSCC)}:444--45


\bibitem{Cao_IMS_2017}
Cao N, Nasir SB, Sen S and Raychowdhury A. 2017. In-Sensor Analytics and Energy-Aware Self-Optimization in a Wireless Sensor Node. {\it IEEE MTT-S International Microwave Symposium (IMS)}:200-03

\bibitem{Chatterjee_TVLSI_2019} 
Chatterjee B, Panda P, Maity S, Biswas A, Roy K, Sen S. 2019. Exploiting Inherent Error Resiliency of Deep Neural Networks to Achieve Extreme Energy Efficiency Through Mixed-Signal Neurons. {\it IEEE Transactions on Very Large Scale Integration (VLSI) Systems} 27(6):1365--77

\bibitem{Chatterjee_ASPDAC_2021}
Chatterjee B, Sen S. 2021. Energy-Efficient Deep Neural Networks with Mixed-Signal Neurons and Dense-Local and Sparse-Global Connectivity : (Invited Paper). {\it 2021 26th Asia and South Pacific Design Automation Conference (ASP-DAC)}:297--304.

\bibitem{Bennett_LNDR_2003}
Bennett CH. 2003. Notes on Landauer's principle, reversible computation, and Maxwell's Demon. {\it Studies in History and Philosophy of Modern Physics} 34(3):501--10

\bibitem{Friis_IRE_1946}
Friis HT. 1946. A Note on a Simple Transmission Formula. {\it Proceedings of the IRE} 34(5):254--56

\bibitem{Salazar_JSSC_2016}
Salazar C, Cathelin A, Kaiser A, Rabaey JM. 2016. A 2.4 GHz Interferer-Resilient Wake-Up Receiver Using A Dual-IF Multi-Stage N-Path Architecture. {\it IEEE Journal of Solid-State Circuits} 51(9):2091--2105

\bibitem{Mohseni_TBioCAS_2015} 
Ebrazeh A, Mohseni P. 2015. 30 pJ/b, 67 Mbps, Centimeter-to-Meter Range Data Telemetry With an IR-UWB Wireless Link. {\it IEEE Transactions on Biomedical Circuits and Systems} 9(3):362--69

\bibitem{Sarpeshkar_AD_1998}
Sarpeshkar R. 1998. Analog Versus Digital: Extrapolating from Electronics to Neurobiology. {\it Neural Computation} 10(7):1601--38

\bibitem{Sodagar_JSSC_2009}
Sodagar AM, Perlin GE, Yao Y, Najafi K, Wise KD. 2009. An Implantable 64-Channel Wireless Microsystem for Single-Unit Neural Recording. {\it IEEE Journal of Solid-State Circuits} 44(9) 2591--2604

\bibitem{Sodagar_TBioCAS_2014}
Hosseini-Nejad H, Jannesari A,  Sodagar AM. 2014. Data Compression in Brain-Machine/Computer Interfaces Based on the Walsh–Hadamard Transform. {\it IEEE Transactions on Biomedical Circuits and Systems} 8(1):129--37

\bibitem{Mich_ISSCC_2020}
Lim J, Moon E, Barrow M, Nason SR, Patel PR, Patil PG, Oh S, Lee I, Kim HS, Sylvester D, Blaauw D, Chestek CA, Phillips J, Jang T. 2020. A 0.19×0.17mm$^2$ Wireless Neural Recording IC for Motor Prediction with Near-Infrared-Based Power and Data Telemetry. {\it 2020 IEEE International Solid- State Circuits Conference - (ISSCC)}: 416-18

\bibitem{Schwartz_ARNeuro_2004} 
Schwartz AB. 2004. Cortical Neural Prostheses. {\it Annual Review Neuroscience} 27:487–-507

\bibitem{Schwartz_TNSRE_2003}
Taylor DM, Tillery IH, Schwartz AB. 2003. Information Conveyed through Brain-Control: Cursor versus Robot. {\it IEEE Transactions on Neural Systems and Rehabilitation Engineering} 11:195–-99

\bibitem{Wise_JSSC_2005}
Olsson RH, Wise KD. 2005. A Three-dimensional Neural Recording
Microsystem with Implantable Data Compression Circuitry. {\it IEEE Journal of Solid-State Circuits} 40(12):2796–-2804
 
\bibitem{Harrison_JSSC_2007}
Harrison R, Watkins PT, Kier RJ, Lovejoy RO, Black DJ, Greger B, Solzbacher F. 2007. A Low-power Integrated Circuit for a Wireless 100-electrode Neural Recording System. {\it IEEE Journal Solid-State Circuits} 42(1):123–-33

\bibitem{SPIKE_OVQ_2011}
Craciun S, Cheney D, Gugel K, Sanchez JC, Principe JC. 2011. Wireless Transmission of Neural Signals using Entropy and Mutual Information Compression. {\it IEEE Transactions on Neural Systems and Rehabilitation Engineering} 19(1):35–44


\bibitem{SPIKE_CS_2012} 
Shoaib M, Jha NK, Verma N. 2012. Enabling Advanced Inference on Sensor Nodes through the Direct Use of Compressively-Sensed Signals. {\it 2012 IEEE Design, Automation \& TEST in Europe Conference}: 437–-43.

\bibitem{SPIKE_Allstot_CS_2012}
Dixon AMR, Allstot EG, Gangopadhyay D, Allstot DJ. 2012. Compressed Sensing System Considerations for ECG and EMG Wireless Biosensors. {\it IEEE Transactions on Biomedical Circuits and Systems} 6(2):156–-66

\bibitem{Mohseni_CS_2018}
Zamani H, Bahrami HR, Chalwadi P, Garris PA, Mohseni P. 2018. C–FSCV: Compressive Fast-Scan Cyclic Voltammetry for Brain Dopamine Recording. {\it IEEE Transactions on Neural Systems and Rehabilitation Engineering} 26(1):51--59

\bibitem{SPIKE_DWT_2007}
Kamboh AM, Raetz M, Oweiss KG, Mason A. 2007. Area-Power Efficient VLSI Implementation of Multichannel DWT for Data Compression in Implantable Neuroprosthetics. {\it IEEE Transactions on Biomedical Circuits and Systems} 1(2):128–-35

\bibitem{SPIKE_DWT_2010}
Yang Y, Kamboh AM, Mason AJ. 2010. Adaptive Threshold Spike Detection using Stationary Wavelet Transform for Neural Recording Implants. {\it 2010 IEEE Biomedical Circuits and Systems Conference (BioCAS)}: 9–-12

\bibitem{SBP_2016} 
Irwin ZT, Thompson DE, Schroeder KE, Tat DM, Hassani A, Bullard AJ, Woo SL, Urbanchek MG, Sachs AJ, Cederna PS, Stacey WC, Patil PG, Chestek CA. 2016. Enabling Low-Power, Multi-Modal Neural Interfaces Through a Common, Low-Bandwidth Feature Space. {\it IEEE Transactions on Neural Systems and Rehabilitation Engineering} 24(5):521--31.

\bibitem{Mohseni_Asilomar_2020}
Wang W, Zhu L, Marefat F, Mohseni P, Kilgore K, Najafizadeh L. 2020. Photoplethysmography-Based Blood Pressure Estimation Using Deep Learning. {\it 2020 Asilomar Conference on Signals, Systems, and Computers}: 945--49

\bibitem{Mohseni_EMBC_2021}
Wang W, Mohseni P, Kilgore K, Najafizadeh L. 2021. Cuff-Less Blood Pressure Estimation via
Small Convolutional Neural Networks. {\it 2021 Annual International Conference of the IEEE Engineering in Medicine and Biology Society (EMBC)}




\bibitem{AES_2017} 
Thomson I. 2017. AES-256 Keys Sniffed in Seconds using \$200 Kit a few inches away. https://www.theregister.co.uk/2017/06/23/aes\_256\_cracked\_50\_seconds\_200\_kit/

\bibitem{Das_ISSCC_2020}
Das D, Danial J, Golder A, Modak N, Maity S, Chatterjee B, Seo DH, Chang M, Varna A, Krishnamurthy H, Matthew S, Ghosh S, Raychowdhury A, Sen S. 2020. 27.3 EM and Power SCA-Resilient AES-256 in 65nm CMOS Through $>$350× Current-Domain Signature Attenuation. {\it 2020 IEEE International Solid- State Circuits Conference - (ISSCC)}: 424--26

\bibitem{Ghosh_ISSCC_2021} 
Ghosh A, Das D, Danial J, De V, Ghosh S, Sen S. 2021. An EM/Power SCA-Resilient AES-256 with Synthesizable Signature Attenuation Using Digital-Friendly Current Source and RO-Bleed-Based Integrated Local Feedback and Global Switched-Mode Control. {\it 2021 IEEE International Solid- State Circuits Conference (ISSCC)}: 499--501

\end{thebibliography}
\end{document}